\pdfoutput=1
\documentclass[aps,prb,reprint,groupedaddress,showpacs,amsmath,amssymb]{revtex4-1}
\usepackage{graphicx}
\usepackage{dcolumn}
\usepackage{bm}
\usepackage{float}
\usepackage{natbib}
\begin{document}
\title{Infrared spectroscopy of silicon for applications in astronomy}
\author{B. Uzakbaiuly}
\author{D. B. Tanner}
\affiliation{Department  of  Physics,  University  of  Florida,  Gainesville,  FL  32611-8440,  USA}
\author{J. Ge}
\affiliation{Department  of  Astronomy,  University  of  Florida,  Gainesville,  FL  32611-8440,  USA}
\author{J. Degallaix}
\affiliation{Albert-Einstein-Institut, Max-Planck-Institut fur Gravitationsphysik, D-30167 Hannover, Germany}%
\author{A. S. Markosyan}
\affiliation{Stanford University, Stanford, CA 94305, USA}
\date{\today}
\begin{abstract}
This work focuses on the characterization of various bulk silicon (Si) samples
using Fourier Transform InfraRed (FTIR) and grating spectrometers in order to get
them suitable for applications in astronomy. Different samples at different impurity
concentrations were characterized by measuring their transmittance in the infrared
region. Various lines due to residual impurity absorption were identified and temperature dependence of impurity absorption is presented. Concentrations of doped samples ($\rho$ $\approx$ 0.2 - 25000 $\Omega$cm) were determined from impurity absorption at low temperatures and from Drude free carrier absorption at 300K.    
\end{abstract}
\maketitle

\section{Introduction}

This work is based upon characterization of silicon (Si) for astrophysical applications. Two proposed applications motivate this work. One is the use of silicon immersion gratings (SIGs) in future infrared spectrometers for satellite telescopes. Immersion grating technology has emerged as an alternative for conventional echelle grating technology in applications where size and weight are important\cite{ge-2012b-v2, Marsh-2007, Ohmori-2001, Wiedemann-1993, yin-2004-gig}. Instead of having incident light reflected and diffracted from the front surface of a conventional diffraction grating, the immersion grating prior light is transmitted into the medium. Diffraction occurs at the back surface. The wavelength of the light inside of the grating is reduced by refractive index of the medium. So the resolution and dispersion are going to be increased by same amount. Since Si has refractive index of 3.4, SIGs offer 3.4 times gain in spectral resolution over conventional echelle gratings of the same length and blaze angle\cite{ge-2014-berik}. This gain can also be viewed as a ~10 times reduction in the instrument size, meaning that the instrument would weight less and be more suitable for space or airborne applications. SIGs have silicon (Si) as the host material, relying on its high transparency in much of the infrared region.

Si is transparent below the multiphonon absorption for far infrared (10 - 600 cm$^{-1}$) use as well as in the near infrared (2000 - 9000 cm$^{-1}$). However, in the far infrared, narrow lines due to residual impurities appear at low temperatures. These will play a significant role in the performance of SIGs and characterization of high purity of Si is required for optimal operation conditions.

A second application is related to incorporating Si in next generation telescopes and detectors. Cryogenic operations are expected to produce substantial improvements in the performance of gravitational-wave detectors. The required position accuracy of the test masses in future detectors is in the range of $10^{-21}$~m or better.\cite{Hild11cqg-v2, Hammond16, Evans16-v2} Achieving this precision will require  improvements over current technologies in many areas, including the massive substrates for the interferometer test masses. KAGRA,\cite{kuroda1, kuroda2, Somiya12cqg, Dooley16, Flaminio16, Tomaru16} a 3 km underground gravitational-wave detector   and the proposed European Einstein Telescope (ET)\cite{Hild11cqg-v2, Punturo10cqg-v2, Puppo11grg, Sathyaprakash12cqg-v2, Adya16}  are exploring cryogenic operations of gravitational-wave detectors, but significant technological hurdles remain. It is important first to select a mirror substrate compatible with cryogenic operation. The material must have good optical qualities, high thermal conductivity, low thermoelastic coefficient, and be available as large, high quality single crystals. KAGRA has selected sapphire for the cryogenic test masses\cite{Somiya12cqg, Hirose14prd} whereas the ET design is based on the use of silicon.\cite{Schnabel10jp} 

Owing to its technological importance, silicon has been studied in great detail.  Crystals as large as 40 cm diameter, 1.1~m in length and weighing more than 425 kg have been produced.\cite{nolimits}  The semiconductor industry routinely cuts and polishes these crystals into wafers up to $30\, ${cm} in diameter; development of $45\, ${cm} diameter materials is continuing,\cite{lu&kimbel-2011} with research  being conducted in the US,  Europe, and Japan.  Silicon has  already met the mass goals for the test masses of   future gravitational-wave detectors. Suspensions for a cryogenic silicon test masses have been designed and studies are in progress.\cite{Shapiro16}

Test masses require low absorption coefficients to provide high transmissive throughput and minimize thermal distortions. Homogeneous optical properties are also required. Optical absorption in substrates results in significant changes in the radii of curvature of the mirror surfaces (due to thermal expansion) and the focal power of the substrates (due to thermal lensing), changing the fundamental arm cavity mode structure.\cite{quetschke2006} Silicon operating at cryogenic temperatures is a particularly strong candidate for third-generation detector test masses. In some experiments, absorption falls below $10^{-7}$ cm$^{-1}$, 10 times lower than absorption in fused silica at $1064\,${nm}.\cite{green95,bruckner10,Steinlechner12cqg} However, other experiments show considerably higher absorption.\cite{Degallaix14cqg}

\subsection{Hydrogen-Like Impurity States in Si}

There have been numerous experimental and theoretical studies to investigate donor \cite{kohnandluttinger-1955, kohn-1955, aggarwal&ramdas-1965, jagannath-1981} or acceptor \cite{Hrostowski-1958, white-1967-fieldsinsi:B, kittel-1954, ramdas&rodriguez-1981-bigrev} defects appearing in Si lattice. Most of them use a picture where an electron or a hole at the impurity is assumed to be in the screened Coulomb potential of the ion core. This screening is due to the dielectric constant of the host atom, Si. The effective mass of the carriers in the Si band structure plays a role as well. 	
The problem of a group VI/IV donor/acceptor atoms in group V Si is very similar to the hydrogen atom problem in quantum mechanics because the extra electron/hole of the donor can be assumed to be influenced by the Coulomb potential screened by a dielectric constant $\epsilon$. The Hamiltonian can be formally written as
\begin{equation}
H_{EM}=-\frac{\hbar^{2}}{2m^{*}}\nabla^{2} - \frac{e^{2}}{4\pi\varepsilon_{0}\varepsilon_{s}r}.
\end{equation}
The energies of this particle would be just the Rydberg energies of hydrogen atom modified by the effective mass (EM) $\left(m_{e}^{*}\right)$ and dielectric constant $\epsilon$ if the conduction/valence band were described by simple quadratic formula. However, the conduction band minimum, as well as the valence band maximum, is complex which requires a rigorous approach to solve for donor/acceptor excited states\cite{faulkner-1969,baldereschi&lipari-1973}.

The conduction band minimum which is located close to the X symmetry point of the Brillouin zone has six symmetry points and the constant energy surfaces in k-space can be described by prolate ellipsoids of revolution\cite{kohnandluttinger-1955}. This anisotropy of constant energy surfaces causes the effective mass to be anisotropic: a longitudinal mass along the main axis of the ellipsoid $\left(m_{\Vert}=0.92m\right)$ and a transverse mass perpendicular to the main axis $\left(m_{\bot}=0.19m\right)$. This anisotropy also plays a factor in electical conductivity effective mass.
The valence band maximum for Si crystal is at the $\Gamma$ point and has two types of energy dispersions. The one with a smaller energy dispersion has a larger mass $m_{hh}=0.54m$, and the one with a higher energy dispersion has a lower mass $m_{lh}=0.15m$. These are fourfold degenerate $p_{3/2}$ like states whose constant energy surfaces are warped spheres in k-space. There are also twofold degenrate $p_{1/2}$ states which are seperated from the former by an energy of $\Delta_{so}$= 0.042 eV \cite{ramdas&rodriguez-1981-bigrev}. Their constant energy surfaces are spheres in k-space corresponding to a mass of $m_{so}=0.24m$. This complex valence band of Si will result in a complex Hamiltonian \cite{luttinger-1956} and will affect the electronic transitions of acceptor states. 
In this paper, the assignment of phosphorus (P) excited states in Si are derived from the theoretical results of Faulkner\cite{faulkner-1969} and the assigment of boron (B) excited states in Si are derived from the results of Steger $et$ $al$ \cite{steger-2009-shallowest}. 

\subsection{Oscillator Strength of Transitions from ground state}\label{sec-OS}

The oscillator strengths of the various transition lines for P donor impurity in Si have been calculated by Clauws $et \ al.$ \cite{clauwsetal-1988-os}. They assumed a Hamiltonian  with a potential defined by
\begin{equation}
\frac{2}{r}\left[1+\left(\epsilon_{\infty}-1\right)e^{-\alpha'r}\right],
\label{imp-potential}
\end{equation}
where $\alpha'$ is a phenomenological parameter which changes according to the ground-state energy value. The wavefunctions of the above Hamiltonian are used to find the oscillator strength
\begin{equation}
f_{a\rightarrow b}=\frac{2m_c^{*}}{\hbar^{2}}\left(E_{b}-E_{a}\right)\left|\boldsymbol{\hat{e}\cdot r_{ba}}\right|^{2},
\end{equation}
where $\boldsymbol{\hat{e}}$ is the unit polarization vector of the radiation and $\boldsymbol{r_{ba}}$ is the dipole matrix element. This is an electric-dipole transition between state $a$ with energy $E_{a}$ and state $b$ with energy $E_{b}$  $\left(E_{a}<E_{b}\right)$. In addition, the donor effective mass is given by equation

\begin{equation}
m^{*}=\frac{3m_{\Vert}m_{\bot}}{2m_{\Vert}+m_{\bot}}\nonumber.
\end{equation}

The transition probabilities are normalized making $\sum_{b}f_{a\rightarrow b}=1$. Calculated oscillator strengths are the transition strengths from $1s\left(A_{1}\right)$ ground state to higher states which is either to odd parity $m_l=0$ or $m_l=\pm1$. 

Another authors (Ref. \onlinecite{beinikhes&kogan-1987-os}) also determined the oscillator strengths of the ground-state to higher-state transitions. Using the wave function for the Hamiltonian and a non-variational method, they arrive at the solution for the oscillator strengths
\begin{eqnarray}
f\left(nP_{0}\right)&=&\frac{2\zeta}{2+\zeta}\left[\tilde{E}\left(nP_{0}\right)-\tilde{E_{0}}\right]\times\nonumber\\
&&\times\left|\sum_{L\geqslant1}\int_{0}^{\infty}d\tilde{r}\tilde{r}^{3}F_{L}^{\left(0\right)}\left(\tilde{r}\right)\bar{R}_{L}^{\left(0\right)}\left(\tilde{r}\right)\right|^{2}\nonumber\\
f\left(nP_{\pm}\right)&=&\frac{4}{2+\zeta}\left[\tilde{E}\left(nP_{\pm}\right)-\tilde{E_{0}}\right]\times\nonumber\\
&&\times\left|\sum_{L\geqslant1}\int_{0}^{\infty}d\tilde{r}\tilde{r}^{3}F_{L}^{\left(1\right)}\left(\tilde{r}\right)\bar{R}_{L}^{\left(1\right)}\left(\tilde{r}\right)\right|^{2},\nonumber\\
\end{eqnarray}
where $R_{L}\left(\tilde{r}\right)$ and $F_{L}^{\left(M\right)}\left(\tilde{r}\right)$ stand for the radial wave function of the initial ground and final state respectively. 

\begin{table}
\caption{Calculated oscillator strength of shallow P donor transitions from the $1S\left(A_{1}\right)$ state in Si\label{tab:Calculated-OS-P}}
\begin{ruledtabular}
\begin{tabular}{c c c c}
Final state & Energy of transition for P$^{\cite{steger-2009-shallowest}}$&OS$^{\cite{clauwsetal-1988-os}}$&OS$^{\cite{beinikhes&kogan-1987-os}}$\\
&meV (cm$^{-1}$) \\
\colrule
$2P_{0}$   & 34.109(275.108) & 0.0312 & 0.0313\\
$2P_{\pm}$ & 39.175(315.966) & 0.1325 & 0.133\\
$3P_{0}$   & 40.104(323.46)  & 0.0064 & 0.00644\\
$3P_{\pm}$ & 42.458(342.45)  & 0.030 & 0.0304\\
$4P_{\pm}$ & 43.388(349.95)  & 0.0108 & 0.0108\\
$5P_{\pm}$ & 44.119(355.84)  & 0.0088 & 0.00909\\
\end{tabular}
\end{ruledtabular}
\label{theory-OS-P}
\end{table}

Table \ref{tab:Calculated-OS-P} lists the oscillator strengths for selected transition lines of phosphorus (P) impurity in the Si lattice as calculated by Clauws $et$ $al.$\cite{clauwsetal-1988-os} and Beinikhes and Kogan\cite{beinikhes&kogan-1987-os}. The overall pictures seem to agree with each other. As can be noted, the $1s\left(A_{1}\right)\rightarrow nP_{\pm}$ transitions are stronger than $1s\left(A_{1}\right)\rightarrow nP_{0}$ ones. This is expected when considering multi-valley degeneracy and arbitrary choice of polarization vector.
As for boron impurity in Si, the oscillator strengths have been calculated by Buczko and Bassani\cite{Buczko&bassani-1992-OSofSi:B}, and Pajot $et$ $al.$\cite{pajotetal-1992-OSofSi:B}. Buczko assumed same Couloumb potential of the negatively charged accepter center as in Eq. \ref{imp-potential} and used variational method to arrive at the oscillator strength derived from the Luttinger Hamiltonian\cite{luttinger-1956}. The oscillator strength for optical transitions, which should obey the sum rule, is

\begin{equation}
f_{0f}=\frac{2m_v^{*}}{\gamma_{1}\hbar^{2}}\frac{1}{g_{0}}\left|E_{f}-E_{0}\right|\sum_{i,j}|\left\langle \Phi_{0,i}|z|\Phi_{f,j}\right\rangle|^{2}
\label{oscil-si:b},
\end{equation}
where $\gamma_{1}$ is Luttinger parameter and $m_v^{*}$ is the conductivity effective mass for holes
\begin{equation}
\frac{1}{m_v^*}=\frac{m_{hh}^{1/2}+m_{lh}^{1/2}}{m_{hh}^{3/2}+m_{lh}^{3/2}}.\nonumber
\end{equation}
Calculated oscillator strengths are transitions from $1\Gamma_{8}^{+}$ ground state to excited states. 

Pajot $et\ al.$ has calculated the oscillator strengths of transitions from ground state using a non variational method. Assuming acceptor-dependent ground state wavefunction, he arrived at electric dipole transition

\begin{eqnarray}
f(a&\rightarrow&b)=\frac{E_b-E_a}{g_a}\nonumber\\
& &\times\sum_{m,n}|\sum_{JFL}(-1)^{F-F_x}C_{nF_z}^{F_a\Gamma_a}C_{mF_z}^{F_b\Gamma_b}\left\{\begin{array}{ccc}F_a&1&F_b\\-F_z&0&F_z\end{array}\right\}\nonumber\\
& &\times(L_aJ_aF_a||r||L_bJ_bF_b)|^2,
\end{eqnarray}

where $a$ and $b$ denote two discrete states. The oscillator strength is normalized to unity which corresponds to sum of discrete as well as continuous transitions. Table \ref{tab:Calculated-OS-B} gives the OS of transitions from ground state $1\Gamma_{8}^{+}$ to the indicated excited state transition in B doped Si.

\begin{table}
\caption{Calculated oscillator strength of shallow B acceptor transitions from the $1\Gamma_{8}^{+}$ state in Si\label{tab:Calculated-OS-B}}
\begin{ruledtabular}
\begin{tabular}{c c c c}
Final state & Energy of transition for B$^{\cite{steger-2009-shallowest}}$&OS$^{\cite{Buczko&bassani-1992-OSofSi:B}}$&OS$^{\cite{pajotetal-1992-OSofSi:B}}$\\
&meV (cm$^{-1}$) &$\times10^{-4}$ & $\times10^{-4}$ \\
\colrule
$1\Gamma_{8}^{-}$   & 30.3694(244.946)  & 194 & 177\\

$2\Gamma_{8}^{-}$   & 34.5042(278.295)  & 769 & 640\\

$3\Gamma_{8}^{-}$   & 38.3770(309.531)  & 53.8 & 54\\

$1\Gamma_{6}^{-}$   & 39.5973(319.374)  & 370 & 260\\
$1\Gamma_{7}^{-}$   & 39.6789(320.032)  & 359 & 376\\
$4\Gamma_{8}^{-}$   & 39.9118(321.910)  & 32.1 & 23\\

$5\Gamma_{8}^{-}$   & 41.4748(334.517)  & 5.1 & 17\\

$6\Gamma_{8}^{-}$   & 42.0494(339.151)  & 2.23 & 12\\
$2\Gamma_{7}^{-}$   & 42.1602(340.045)  & 27.4 & 31\\
$7\Gamma_{8}^{-}$   & 42.0494(339.151)  & 3.37 & 3\\

$3\Gamma_{6}^{-}$   & 42.7180(344.544)  & 27.1 & 36\\
$3\Gamma_{7}^{-}$   & 42.7540(344.834)  & 50.6 & 42\\
$8\Gamma_{8}^{-}$   & 42.9361(346.303)  & 6.43 & 5\\

$4\Gamma_{6}^{-}$   & 43.1677(348.171)  & 7.62 & 1.7\\
$4\Gamma_{7}^{-}$   & 43.2683(348.982)  & 10.9 & 0.01\\
$10\Gamma_{8}^{-}$  & 43.305(349.28)    & 0.611 & 1\\
$11\Gamma_{8}^{-}$  & 43.4799(350.689)  & 2.48 & 3.8\\
\end{tabular}
\end{ruledtabular}
\label{theory-OS-B}
\end{table}

\begin{table*}
\caption{Samples measured}
\label{samples-measured}
\begin{ruledtabular}
\begin{tabular}{c c c c c}
Sample ID & Type & Thickness & Resistivity, $\rho$ & Concentration, $n_i$\footnotemark[1] \\
          &      & cm        & $\Omega$ cm$^{-1}$     & cm$^{-3}$ \\
\colrule
si-jge    & n-type/Float zone & 1.29     & N/A  & N/A\\

si-st1    & n-type/Czochralski & 2.02     & 0.17  & 3.9$\times 10^{16}$\\

si-st2    & n-type/Czochralski & 2.02     & 2.8  & 1.6$\times 10^{15}$\\

si-st4    & n-type/Float zone & 2.02     & 808  & 5.2$\times 10^{12}$\\

si-8547    & n-type/Czochralski & 0.63     & 134  & 3.18$\times 10^{13}$\\

si-8548    & p-type/Czochralski & 0.63     & 159  & 8.37$\times 10^{13}$\\

si-8549    & n-type/Float zone & 0.63     & 2950  & 1.39$\times 10^{12}$\\

si-8550    & p-type/Float zone & 0.63     & 115  & 1.16$\times 10^{14}$\\

si-8551/8551*\footnotemark[2]    & n-type/Float zone & 0.63     & 10100  & 4.00$\times 10^{11}$\\

si-8552/8552*\footnotemark[2]    & p-type/Float zone & 0.63     & 28500  & 4.67$\times 10^{11}$\\

\end{tabular}
\end{ruledtabular}
\footnotetext[1]{As determined from resistivity using the ASTM standard\cite{standard-2000-restoconc}}
\footnotetext[2]{Two samples of same type but taken from different parts of the boule}
\end{table*}

\subsection{Lattice Absorption and Vibrational Modes}\label{sec-lattice}

Dipole moments resulting from the lattice vibrations in Si are not first order and, hence, even though there are two Si in unite cell, Si cannot have single phonon absorption. In homopolar diamond-like structures, the absorption of the photon can be described by the simultaneous interaction of two phonons. The first phonon will induce a local charge distortion and the second phonon will simultaneously interact with this charge distortion and produce a temporary dipole moment. This dipole moment can then couple to the incident radiation. The two phonons are usually from two different crystallographic directions \cite{lax&burstein-1955}. In Si, the multiphonon absorptions are infrared active and they can readily be seen in the absorption spectra. Johnson (Ref. \onlinecite{johnson-1958-phononsi}) has measured the optical absorption of Si and has assigned several absorption lines to the two- or three-phonon absorption. 

It is important to note here that the oxygen (O) vibrational modes in Si are also infrared active. Oxygen which enters Si interstitially results in complex of 
Si$_{2}$O type. The fundamental normal modes of vibration of this complex is explained to be the root of some infrared absorption \cite{bosomworth-1970-oxygen&sivibrat}. Several vibrational modes take place. The $\nu_{2}$ vibrational modes happen at 29.3, 37.8, 43.3 and 49.0 cm$^{-1}$ and the $\nu_{3}$ vibrational modes are a stem of 1136 cm$^{-1}$ vibration bands. Temperature-dependent effects arise when transitions occur from thermally populated of the excited states of the $\nu_2$ mode. This absorption happens at 1127.9 cm$^{-1}$.

\section{Experimental Procedures}

Experimental samples, doping levels, along with other properties are presented in Table \ref{samples-measured}. The concentrations of the samples were calculated from given resistivity using the ASTM standard\cite{standard-2000-restoconc}. Specific holders were designed for some of the samples so that there would be good thermal contact with the liquid He LT3B Helitran cryostat. Temperature sensor was attached in close proximity to the holder so as to monitor sample temperature. Transmittance measurements were done by Bruker 113v FTIR spectrometer with Hg source and 4.2K Si bolometer detector pair for far infrared (10-700 cm$^{-1}$) and Globar and DTGS detector pair for mid infrared (400-5000 cm$^{-1}$) ranges. Resolution at low temperatures (10-100 K) was 0.5 cm$^{-1}$ and at higher temperatures (100-300 K) was 4 cm$^{-1}$. The near-infrared spectra (4000-12000 cm$^{-1}$) were measured by using a Perkin Elmer 16U grating spectrometer. In this case, tungsten filament source and photoconductive detector (PbS) pair was used during measurements.

\section{ANALYSIS}
\subsection{Beer-Lambert Law}\label{sec-beer-lamber}

From the transmittance of a sample $\left(T=I_{t}/I_{0}\right)$, assuming no reflection, we can extract the absorption coefficient from the Beer-Lambert law
\begin{equation}
T=\frac{I_{t}}{I_{0}}=e^{-\alpha t} \rightarrow\alpha=-\frac{1}{t}\mathrm{ln}\left(\frac{I_{t}}{I_{0}}\right),
\end{equation}
where $t$ is the thickness of the sample and $\alpha$ is the absorption coefficient which is related to absorption cross section $\sigma_{abs}$ and the concentration of absorbing species $n_i$ through $\alpha=\sigma_{abs} n_i$. 

Since the actual transmission is not that simple because there might be multiple reflection and transmission inside of the sample, more rigorous approach should be taken\cite{tanner-opteff,wooten-1972}. The transmittance of a thick sample which causes the light to be incoherent is
\begin{equation}
T=\frac{\left(1-R_{12}\right)^{2}e^{-\alpha x}}{1-R_{12}^{2}e^{-2\alpha x}},
\label{eq-T-alpha-relation}
\end{equation}
where $R_{12}$ 	is the reflectance at the front surface of the sample
\begin{equation}
R_{12}=\left|\frac{1-\hat{n}}{1+\hat{n}}\right|^{2}=\frac{\left(n-1\right)^{2}+\kappa^{2}}{\left(n+1\right)^{2}+\kappa^{2}}.
\label{R-front-surf}
\end{equation}
The absorption coefficient is related to extinction coefficient through $\alpha=2\omega\kappa/c$. If the absortion of a sample is small enough, that $\kappa\ll n$ then Equations \ref{eq-T-alpha-relation} and \ref{R-front-surf} can be written as
\begin{eqnarray}
&&e^{\alpha x}=\frac{(1-R_{12})^{2}}{T}\left [ 1/2 + \sqrt{1/4+ \frac{R_{12}^{2} T^{2}}{(1-R_{12})^{4}}} \right ]\nonumber\\
&&R_{12}=\frac{\left(n-1\right)^{2}}{\left(n+1\right)^{2}}.
\label{t2a eq}
\end{eqnarray}
Therefore, knowing the refractive index of the sample $n$, $\alpha$ could be extracted from these set of formulas.

Here we can define a more general quantity, integrated absorption or spectral weight $S$, which would be approximately independent of spectral resolution\cite{pajot-2010}
\begin{equation}
S=\int_{\widetilde{v}_{min}}^{\widetilde{v}_{max}} \alpha (\widetilde{v})d{\widetilde{v}} = K_S\:n_i,
\label{eq-intg-abs}
\end{equation}
where $K_S$ is an integrated absorption factor and the integration is over spectral extent of the isolated line. Here the wavenumber is denoted by $\widetilde{v}$. The concentration of a given impurity would be then an integrated calibration factor (i.e., $K_S^{ -1}$) mulitplied by $S$.

\begin{figure*}
\includegraphics[width=1.971\columnwidth]{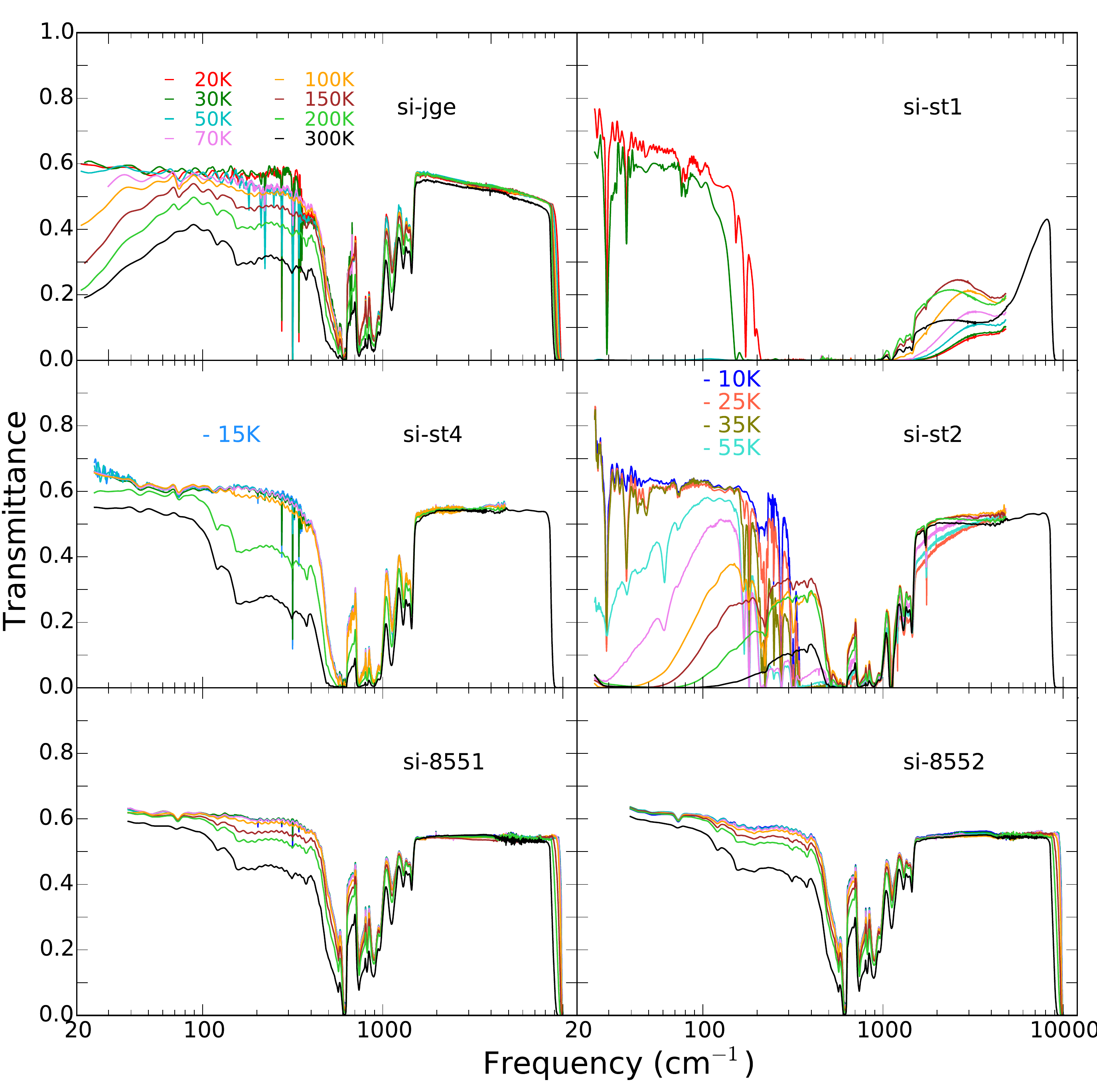}
\caption{\label{all-old-batch}Transmittance of samples measured from 20 or 40 cm$^{-1}$ to 11,000 cm$^{-1}$.}
\end{figure*}
\begin{figure*}
\includegraphics[width=1.971\columnwidth]{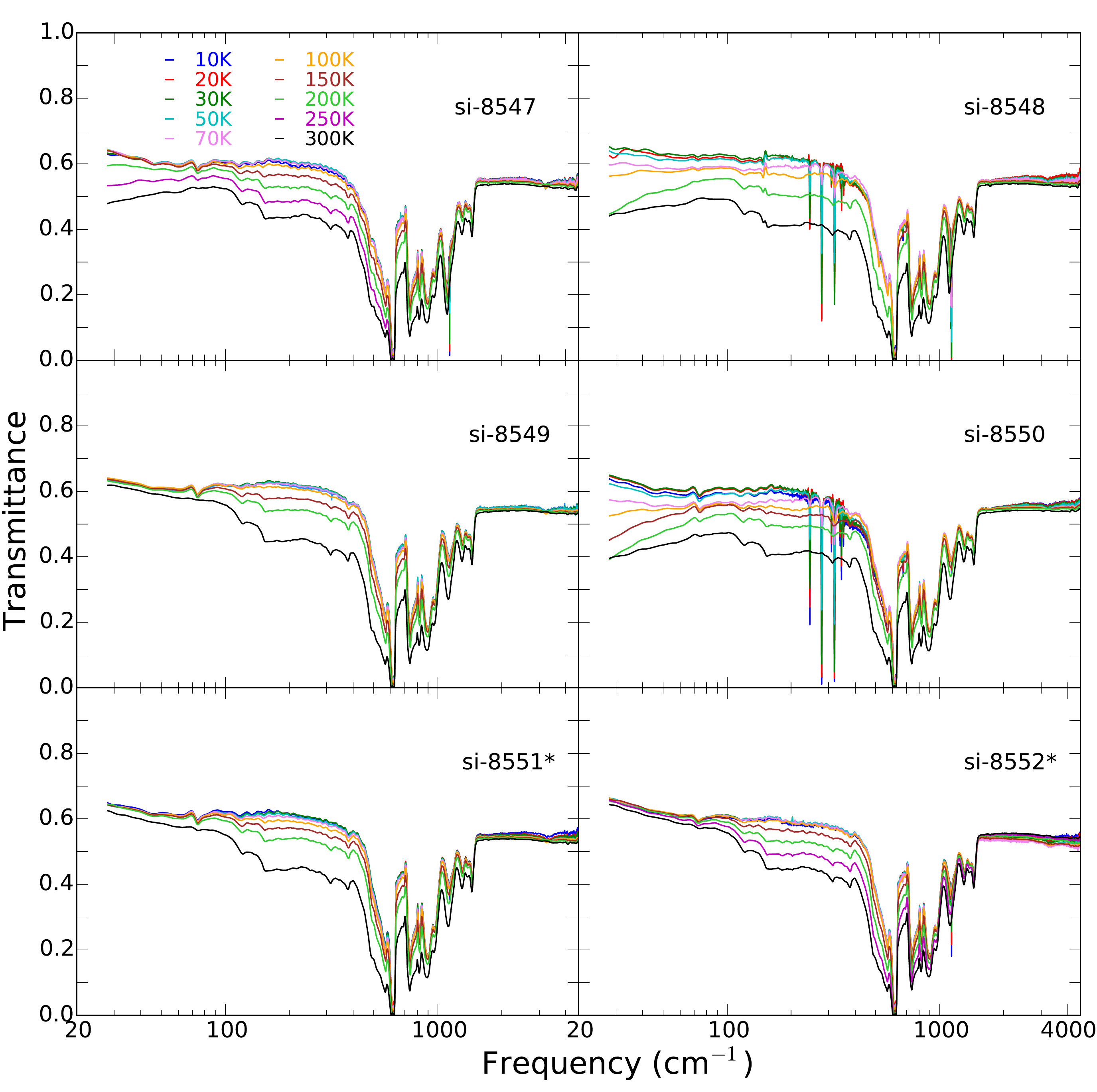}
\caption{\label{all-new-batch}Transmittance of samples from 20 or 30 cm$^{-1}$ to 4500 cm$^{-1}$.}
\end{figure*}

\subsection{Drude-Lorentz models}\label{sec-drude-cond}

Using absorption coefficient ($\alpha$) and assuming constant refractive index $n$, the real part of the optical conductivity ($\sigma_1$) of the sample could be obtained through relation\cite{tanner-opteff}(in cgs units)
\begin{equation}
\alpha=\frac{2\omega\kappa}{c}=\frac{4\pi\sigma_{1}}{nc},
\label{alpha-sigma1 rel}
\end{equation}
where $\kappa$ is extinction coefficient. A Drude model was used to fit the free carrier part of the conductivity spectra. The equations for optical conductivity in the Drude part of the model are\cite{tanner-opteff}
\begin{eqnarray}
&&\sigma=\frac{n_c e^{2}\tau/m}{1-i\omega\tau}\nonumber\\
&&\sigma_{1}=\frac{\sigma_{dc}}{1+\omega^{2}\tau^{2}}\nonumber\\
&&\sigma_{2}=\frac{\omega\tau\sigma_{dc}}{1+\omega^{2}\tau^{2}},
\label{eq-drude-cond}
\end{eqnarray}
where $\tau$ is the scattering time, and $\sigma_{dc}$ is the conductivity when $\omega$=0. Here, it is interesting to look at the Drude conductivity $\sigma_{1}(\omega)$. If we integrate the conductivity from 0 to $\infty$ frequency we get
\begin{equation}
\int_{0}^{\infty}d\omega'\sigma_{1}\left(\omega'\right)=\frac{1}{8}\omega_{p}^{2}=\frac{\pi}{2}\frac{n_c e^{2}}{m}.
\label{eq-sumrule}
\end{equation}
This equation is the sum rule for the Drude conductivity, and is independent of $\tau$. This equation says that an increase in the conductivity at low frequencies, for example, is offset by a decrease at higher frequencies. It should also be noted that the partial sum rules can be applied if the integral is carried out from 0 to some frequency $\omega$. Then only the effective concentration of electrons participating in optical transitions at frequencies lower than 
$\omega$ would be involved. Eq. \ref{eq-sumrule} changes to
\begin{equation}
\frac{\pi}{2}\frac{n_{c_{eff}} e^{2}}{m^{*}}=\int_{0}^{\omega}d\omega'\sigma_{1}\left(\omega'\right).
\label{eq-part-sumrule}
\end{equation}
This equation would be important in fitting for the free-carrier term in the conductivity spectrum.

\section{RESULTS AND DISCUSSION}
\subsection{Transmittance spectra}

We have measured transmission spectra of various Si samples at different doping levels. Some of the samples were measured all the way to the band edge of Si. Others to about half of the bandgap. Fig. \ref{all-old-batch} and \ref{all-new-batch} show the transmittance data for all of these samples at different temperature ranges. In the far infrared, the transmittance of all of the samples rises as the temperature is lowered. Also, sharp absorption lines due to residual impurities start to appear at low temperatures. In the mid infrared range, we see the multiphonon absorption of Si lattice and in the near infrared we see the bandedge of the Si which becomes bigger at low temperatures. The sharp absorption lines are mostly due to ubiquitous impurities in Si, which are P, B, and O. The samples that had P or B in them were analyzed to give approximate concentration. Optical conductivity spectra (namely $\sigma_{1}$), given in subsequent sections, were calculated using Eq. \ref{eq-T-alpha-relation} and \ref{alpha-sigma1 rel}.

\begin{figure}
\includegraphics[width=1\columnwidth]{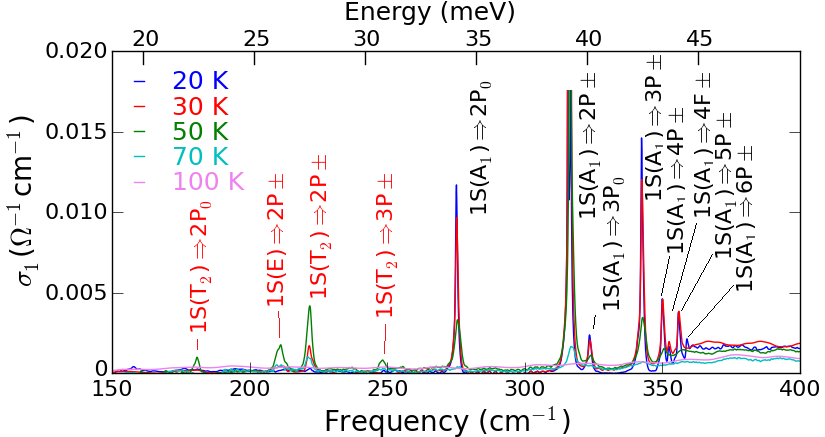}
\caption{\label{farir-si-jge}Optical conductivity $\sigma_1$ of sample si-jge. Lines are attributed to P}
\end{figure}
\begin{figure}
\includegraphics[width=1\columnwidth]{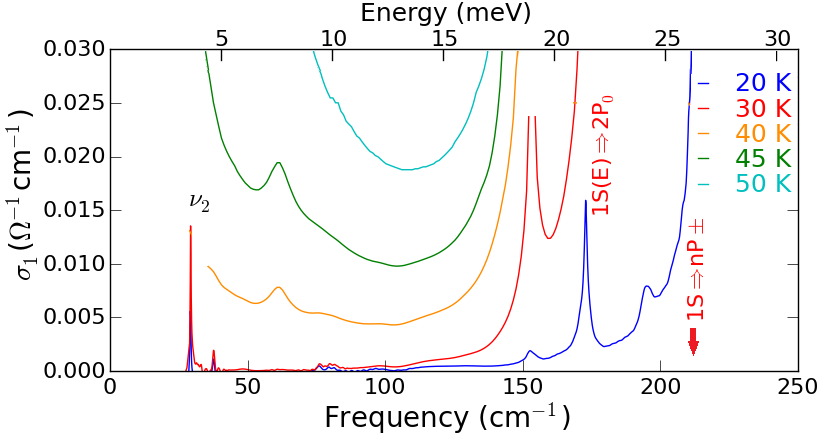}
\caption{\label{farir-si-st1}Optical conductivity $\sigma_1$ of sample si-st1. P-doped}
\end{figure}

\subsection{Transition lines}

\begin{figure}[b]
\centering
\includegraphics[width=1\columnwidth]{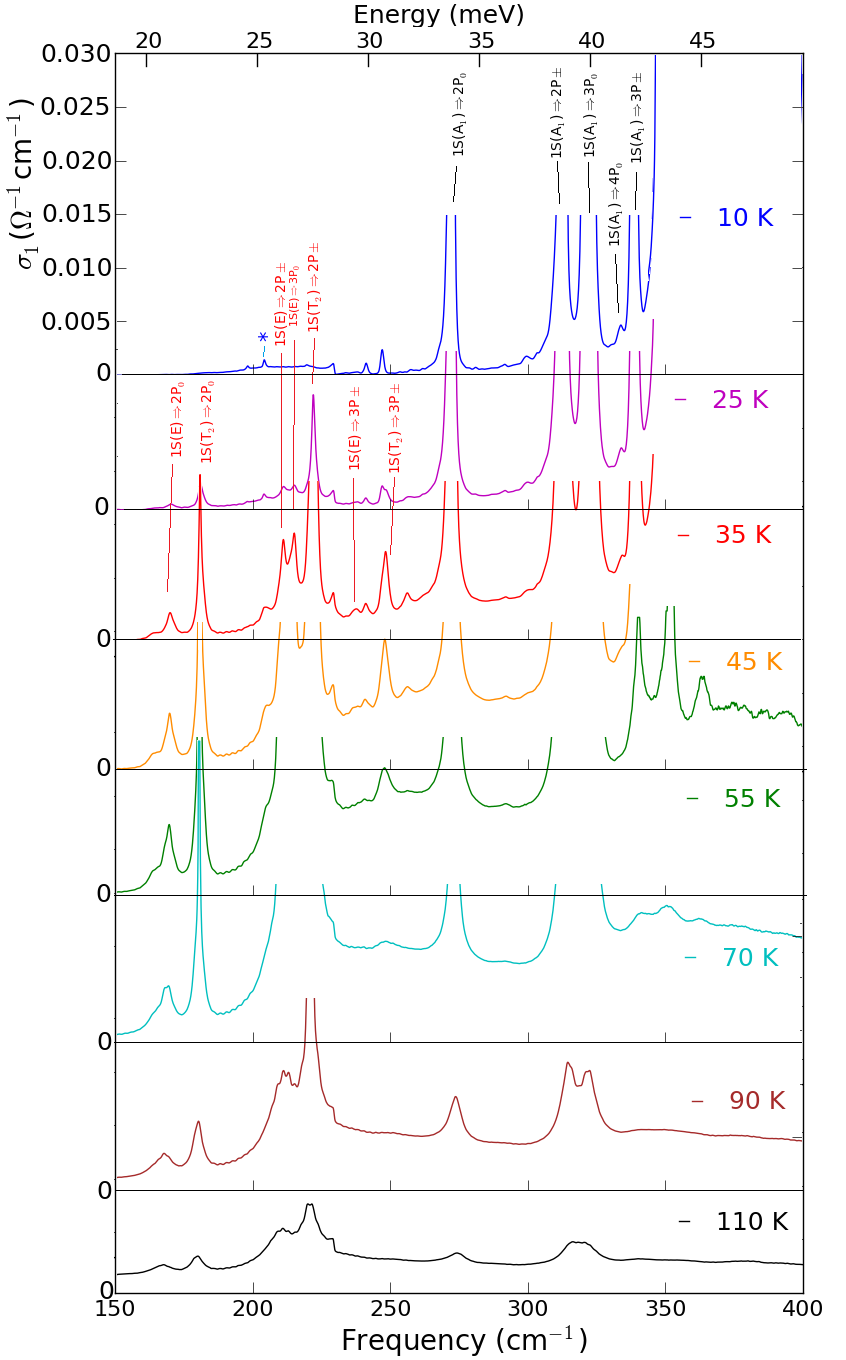}
\caption{\label{farir-si-st2}Optical conductivity $\sigma_1$ of sample si-st2. Sharp lines due to P.}
\end{figure}

\begin{figure}
\includegraphics[width=1\columnwidth]{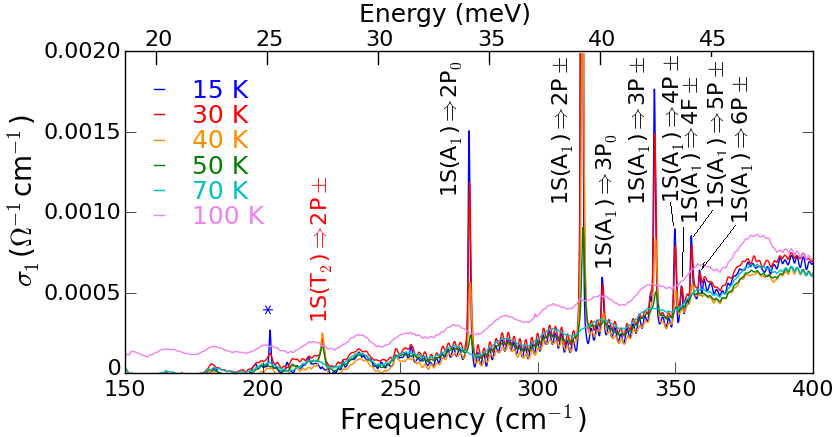}
\caption{\label{farir-si-st4}Optical conductivity $\sigma_1$ of sample si-st4. Sharp lines due to P.}
\end{figure}

\begin{figure}
\includegraphics[width=1\columnwidth]{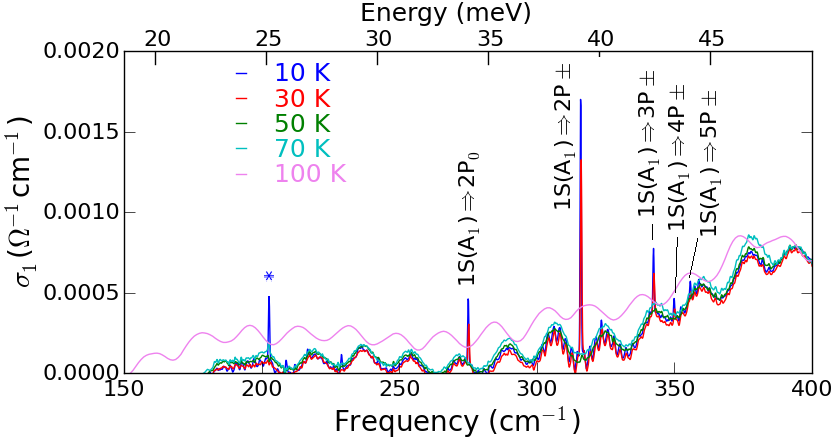}
\caption{\label{farir-si-8551}Optical conductivity $\sigma_1$ of sample si-8551. Sharp lines due to P.}
\end{figure}

\begin{figure}
\includegraphics[width=1\columnwidth]{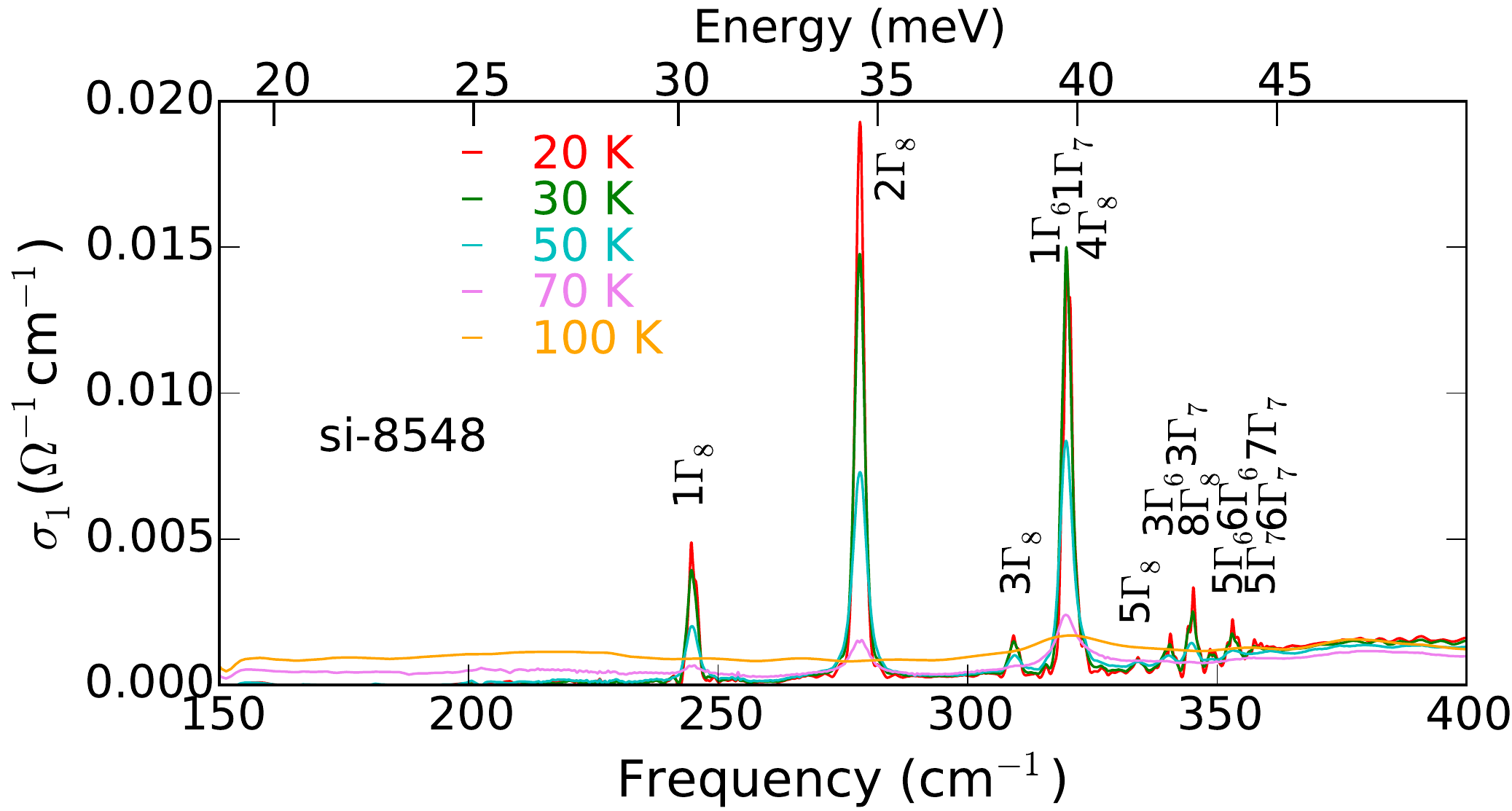}
\caption{\label{farir-si-8548}Optical conductivity $\sigma_1$ of sample si-8548. Sharp lines due to B.}
\end{figure}

\begin{figure}
\includegraphics[width=1\columnwidth]{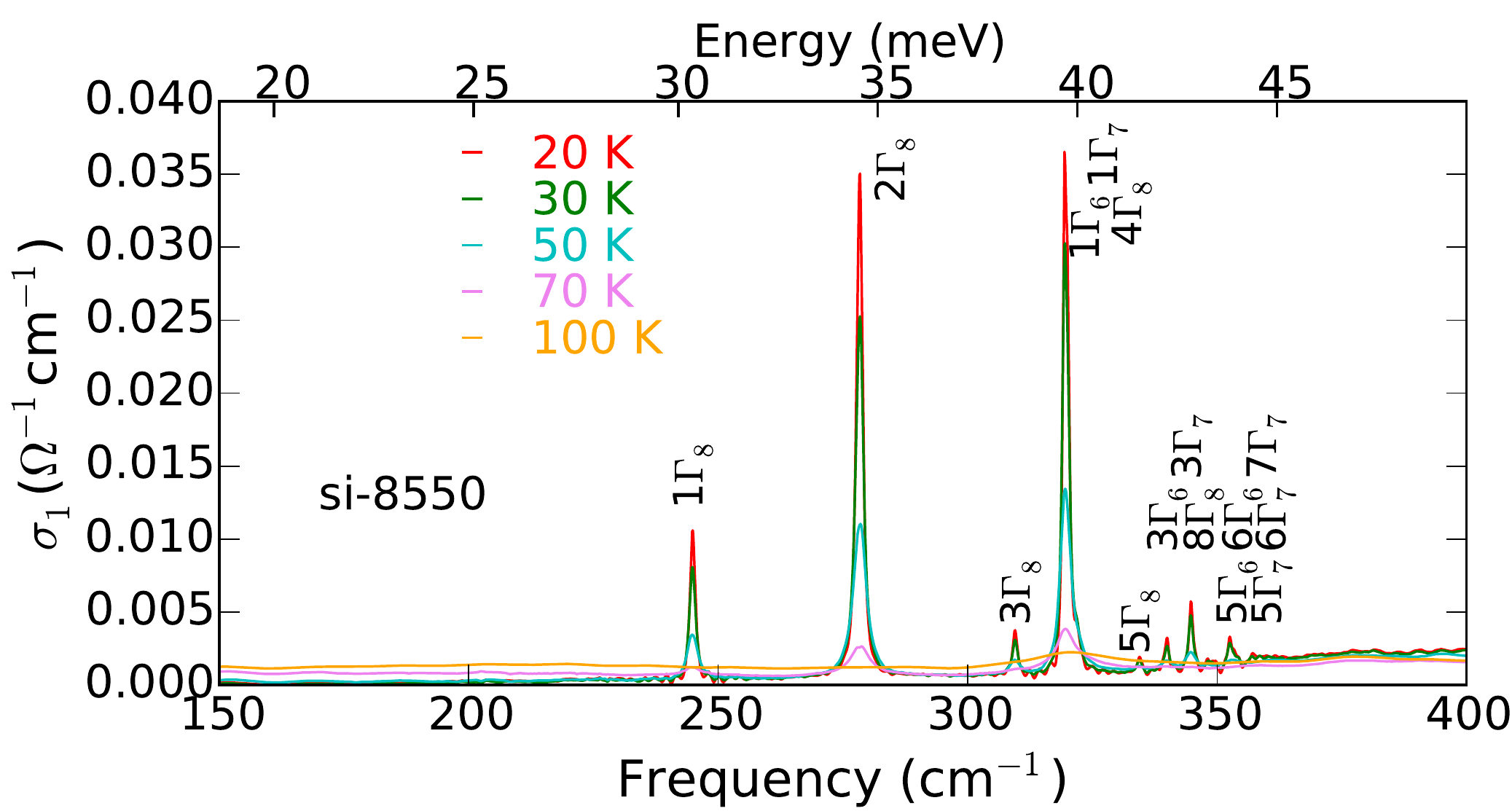}
\caption{\label{farir-si-8550}Optical conductivity $\sigma_1$ of sample si-8550. Sharp lines due to B.}
\end{figure}

First, we discuss the sharp absorption lines that appear in the far infrared due to impurities in the n-type Si samples. Fig. \ref{farir-si-jge} shows the far-infrared optical conductivity spectra of sample si-jge. The various transition lines were identified. They are due to the P impurity whose experimentally determined ionization energy is 45.578 meV \cite{steger-2009-shallowest}. In samples with high impurity level, some of the transition lines had to be truncated because the transmittance is zero within the noise and accuracy of the measurements. For these lines, the absorption coefficient and thus the optical conductivity could not be calculated since when inverting Eq. \ref{t2a eq}, it results in taking natural logarithm of transmittance. Since the $\mathrm{ln}(1/T)$ is undefined, the integrated optical conductivity and the full width at half maximum of the lines are undefined too. Also, we observe that at the intermediate temperatures, there are absorption lines which are below the lowest transition energy of the ground state (i.e., $1S(A_{1})\rightarrow2P_{0}$). These lines are actually from the valley-orbit split ground states of $1S(E)$ and $1S(T_{2})$ because they become thermally populated \cite{aggarwal&ramdas-1965} at the intermediate temperatures. Similar transition lines were observed for sample si-st4 (Fig. \ref{farir-si-st4}) and si-8551 (Fig. \ref{farir-si-8551}). If we now look at si-st2 (Fig. \ref{farir-si-st2}), we can also observe the transitions from both ground state and valley-orbit split ground states to higher states of P impurity. But they are shifted towards the lower energies than for more pure samples. This shiftis actually expected for double donor absorption (i.e., P donor pairs) in a Si sample due to heavier doping\cite{thomas-1981-mitr}. For si-st1, at higher temperatures (100-300 K), we observe some transmission in the mid infrared range, yet we do not observe any transmission in the far infrared range. This opacity is because the sample has a high impurity concentration. Free-carrier 
\clearpage
\hspace{-1em}absorption, coming from electrons ionized from the P atoms, dominates in the far infrared. As temperature decreased below 100 K, some transmission started to happen below 200 cm$^{-1}$ and some absorption lines started to appear. These lines again correspond to P lines in Si, and, since the band of impurities become wide enough, we can no longer see any sharp transition lines corresponding to $1S(A_{1}/E/T_{2})\rightarrow nP_{0/\pm}$ like transition. Instead, we observe a strong absorption edge.
For the p-type Si, sharp absorption lines appear due to the B impurity in Si, whose experimentally determined ionization energy is 45.63 meV \cite{steger-2009-shallowest}. In Fig. \ref{farir-si-8548} and \ref{farir-si-8550}, you can see the far infrared optical conductivity spectra of si-8548 and si-8550, respectively. The various transition lines were identified. They are due to ground state 1$\Gamma_8^+$ to excited state n$\Gamma_n^-$ transitions in boron. These are so called ${p_{3/2}}$ spectra that appear for acceptors in Si. A convolution of some B lines, which are close to each other in the spectrum, are observed due to the limitation in the insturmental resolution. There are no temperature dependent transitions appearing for this case because the spin-orbit-split ground state 1$\Gamma_7^+$, which lies in the band gap, remains depopulated up to room temperature \cite{pajot-2010}.

\subsection{Oscillator strength of hydrogen-like states}

\begin{figure}
\includegraphics[width=1\columnwidth]{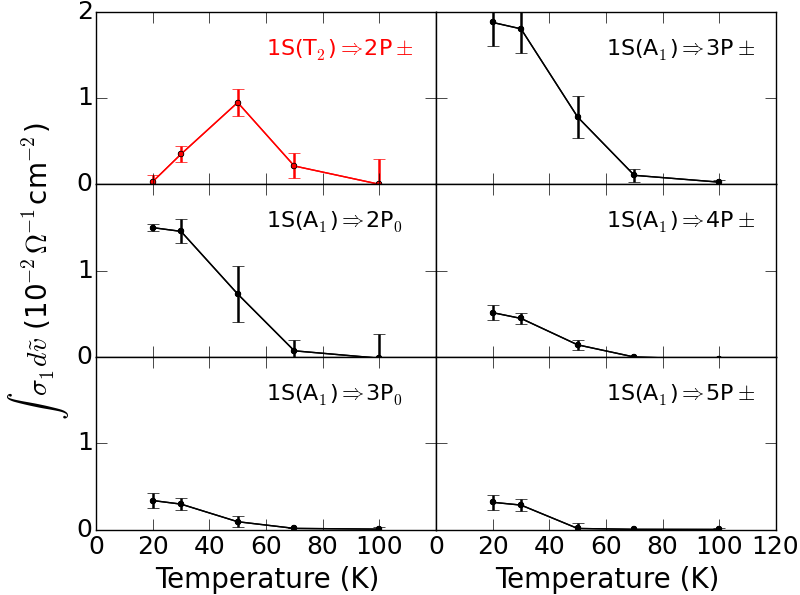}
\caption{\label{stat-si-jge} Integrated optical conductivity vs temperature for some of the lines in sample si-jge. P doped.}
\end{figure}

\begin{figure}
\includegraphics[width=1\columnwidth]{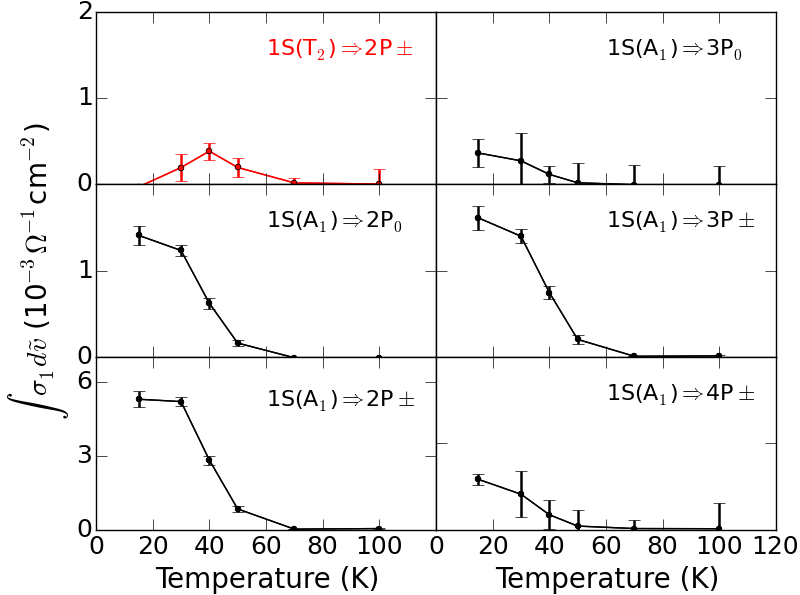}
\caption{\label{stat-si-st4} Integrated optical conductivity vs temperature for some of the lines in sample si-st4. P doped.}
\end{figure}

\begin{figure}
\includegraphics[width=1\columnwidth]{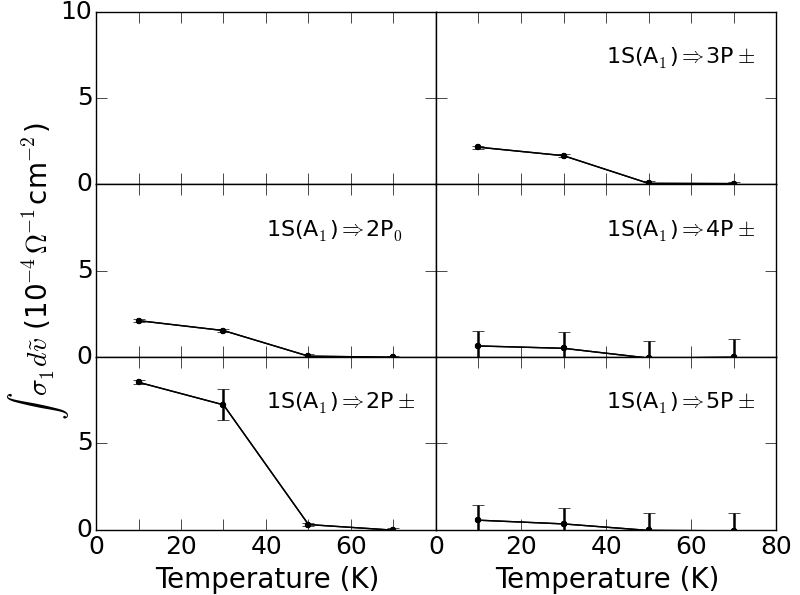}
\caption{\label{stat-si-8551} Integrated optical conductivity vs temperature for some of the lines in sample si-8551. P doped.}
\end{figure}

\begin{figure}
\includegraphics[width=1\columnwidth]{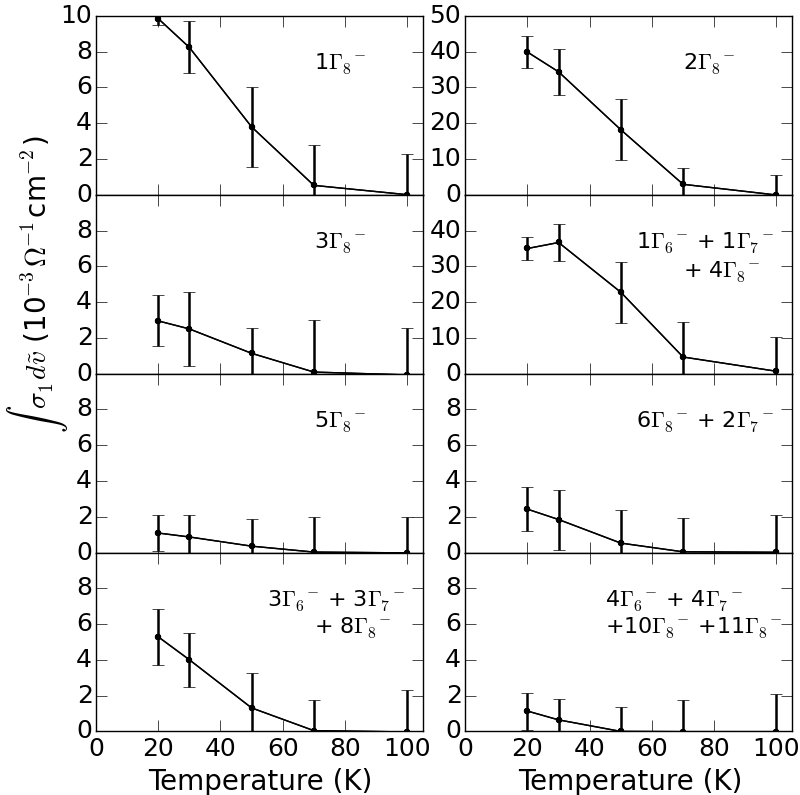}
\caption{\label{stat-si-8548} Integrated optical conductivity vs temperature for some of the lines in sample si-8548. B doped.}
\end{figure}

\begin{figure}
\includegraphics[width=1\columnwidth]{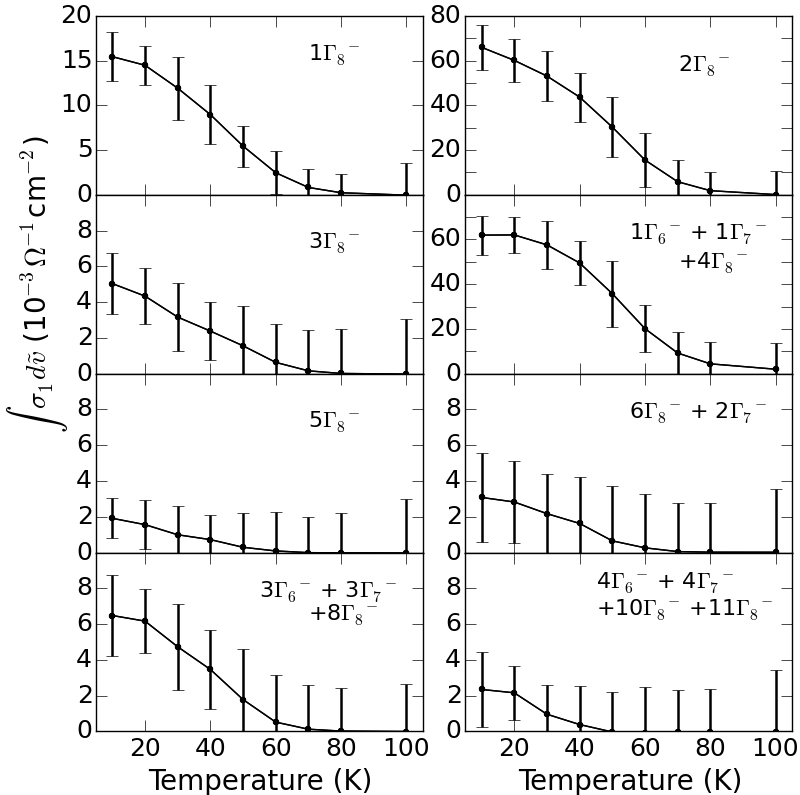}
\caption{\label{stat-si-8550} Integrated optical conductivity vs temperature for some of the lines in sample si-8550. B doped.}
\end{figure}

The temperature dependence of the integrated optical conductivity of transition lines are given in Fig. \ref{stat-si-jge}-\ref{stat-si-8550}. The behavior is as expected. Transition from the ground state to excited states become weak as temperature increases from the lowest temperature, as the ground state becomes depopulated with increasing temperature. Whereas for n-type Si, valley-orbit-split ground state to excited state transition is strong at intermediate temperatures. The integrated optical conductivity for many lines could not be calculated either due to transmission going being zero within the noise or not enough temperature points to see the behavior. For p-type Si, some of the B lines are in close proximity, so the integration is done through spectral extent of several lines which are indicated on the figures.

\subsection{Multiphonon absorption}

\begin{figure}
\includegraphics[width=1\columnwidth]{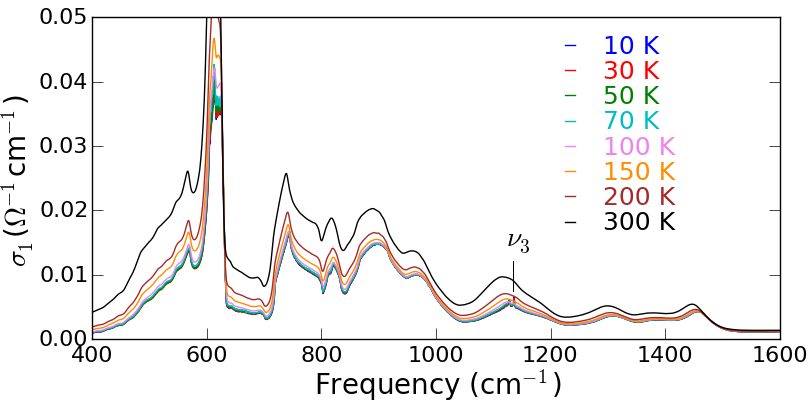}
\caption{\label{multi-si-8552} Multiphonon spectra for sample si-8552}
\end{figure}

\begin{figure}
\includegraphics[width=1\columnwidth]{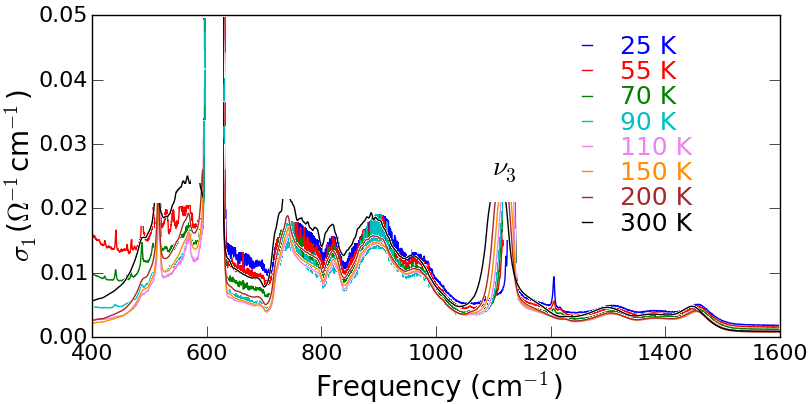}
\caption{\label{multi-si-st2} Multiphonon spectra for sample si-st2}
\end{figure}

Multiphonon absorption happens in the midinfrared range for Si. As discussed in Sec. \ref{sec-lattice}, the Si cannot have a single phonon absorption. The behavior of multiphonon absorption lines with temperature is as expected for all of the samples. The optical conductivity is shown for two of the samples in Fig. \ref{multi-si-8552} and \ref{multi-si-st2} since they represent the behavior of multiphonons for all of the samples. The linewidth narrows as temperature is decreased since lattice vibrations relax. The strongest absorption at around 600 cm$^{-1}$ for some of the samples had to be cut of since the transmittance is zero within the noise and accuracy of the measurements. 
\begin{figure}
\includegraphics[width=1\columnwidth]{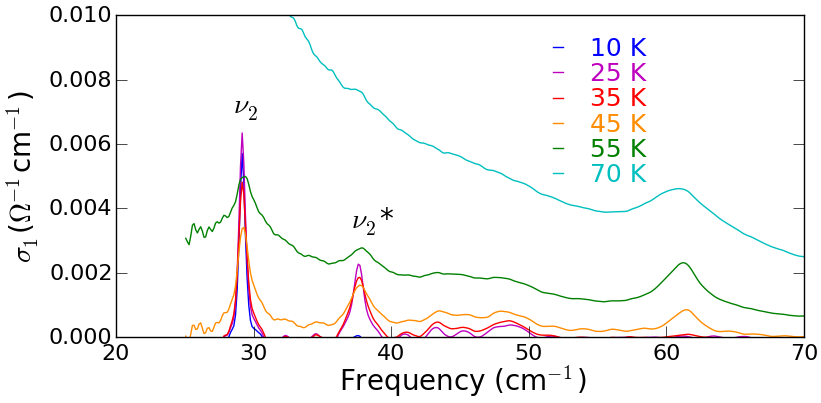}
\caption{\label{v2-si-st2}$\nu_2$ vibrational modes of Si$_2$O vibrations for sample si-st2}
\end{figure}
\begin{figure}
\includegraphics[width=1\columnwidth]{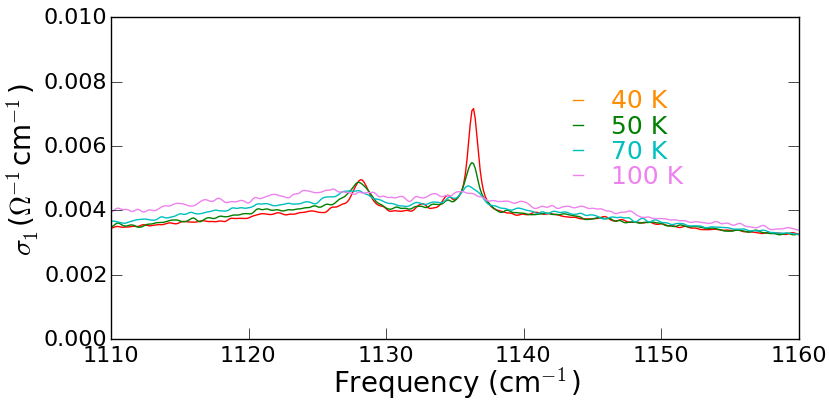}
\caption{\label{v3-si-st4}$\nu_3$ vibrational modes of Si$_2$O vibrations for sample si-st4}
\end{figure}
\begin{figure}
\includegraphics[width=0.988\columnwidth]{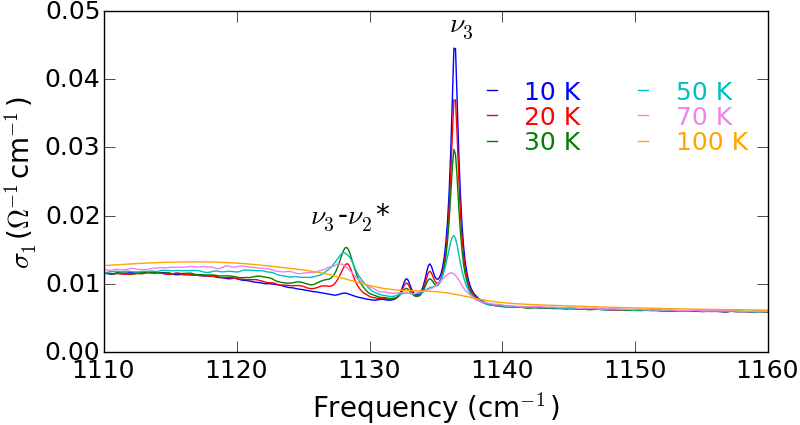}
\caption{\label{v3-si-8547}$\nu_3$ vibrational modes of Si$_2$O vibrations for sample si-8547}
\end{figure}
\begin{figure}
\includegraphics[width=0.988\columnwidth]{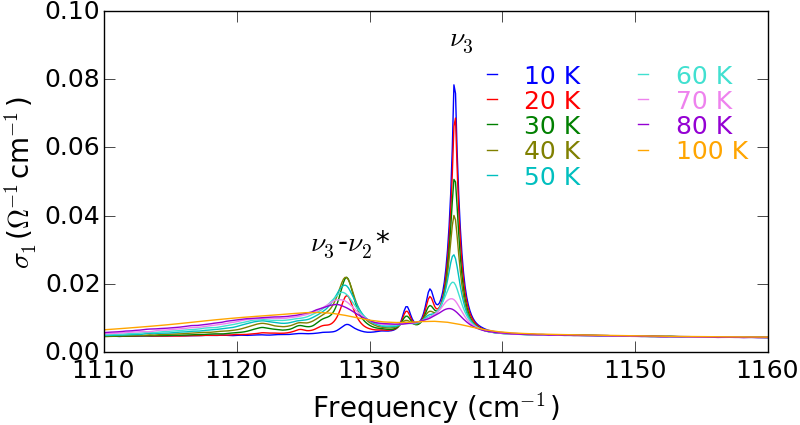}
\caption{\label{v3-si-8548}$\nu_3$ vibrational modes of Si$_2$O vibrations for sample si-8548}
\end{figure}
\begin{figure}
\includegraphics[width=0.988\columnwidth]{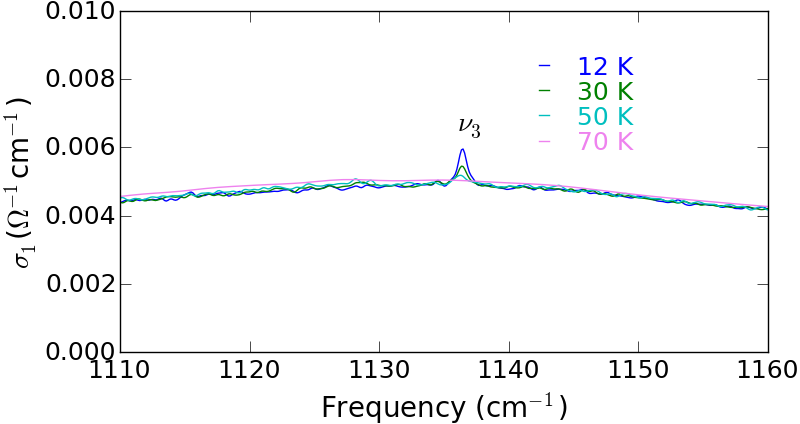}
\caption{\label{v3-si-8549}$\nu_3$ vibrational modes of Si$_2$O vibrations for sample si-8549}
\end{figure}
\begin{figure}
\includegraphics[width=0.988\columnwidth]{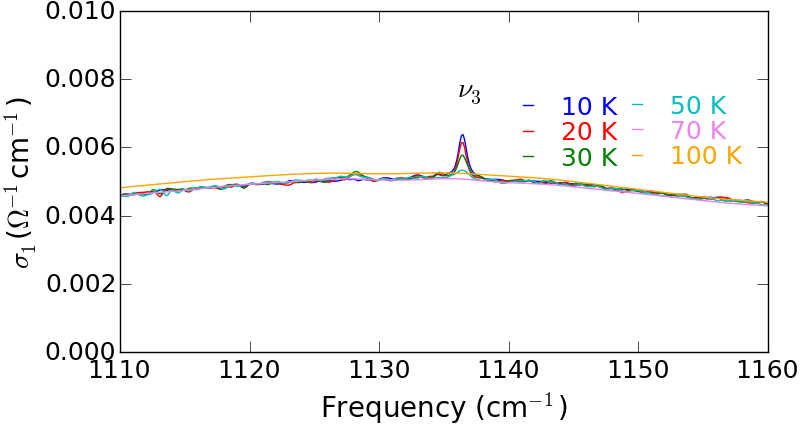}
\caption{\label{v3-si-8550}$\nu_3$ vibrational modes of Si$_2$O vibrations for sample si-8550}
\end{figure}
\clearpage
\begin{figure}
\includegraphics[width=0.98\columnwidth]{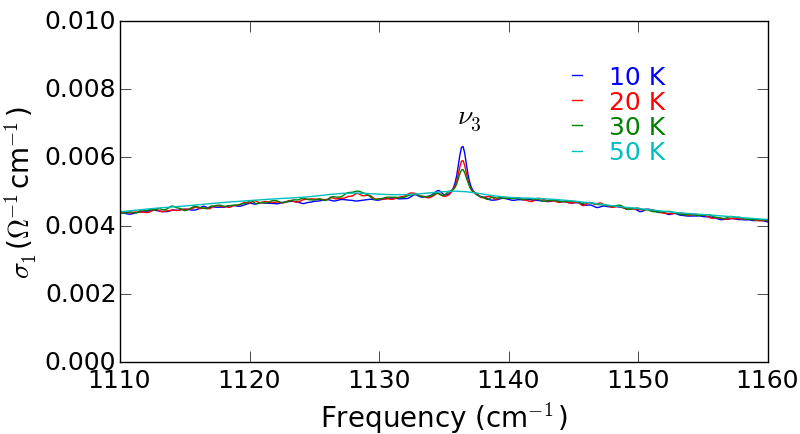}
\caption{\label{v3-si-8551}$\nu_3$ vibrational modes of Si$_2$O vibrations for sample si-8551*}
\end{figure}
\begin{figure}
\includegraphics[width=0.98\columnwidth]{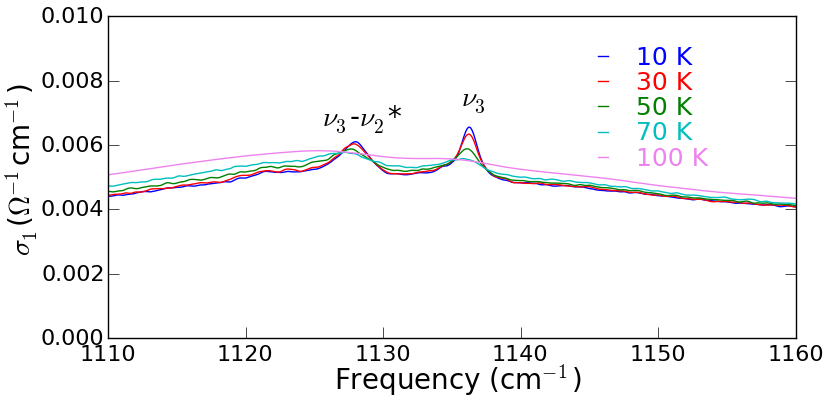}
\caption{\label{v3-si-8552}$\nu_3$ vibrational modes of Si$_2$O vibrations for sample si-8552}
\end{figure}
\begin{figure}
\includegraphics[width=0.98\columnwidth]{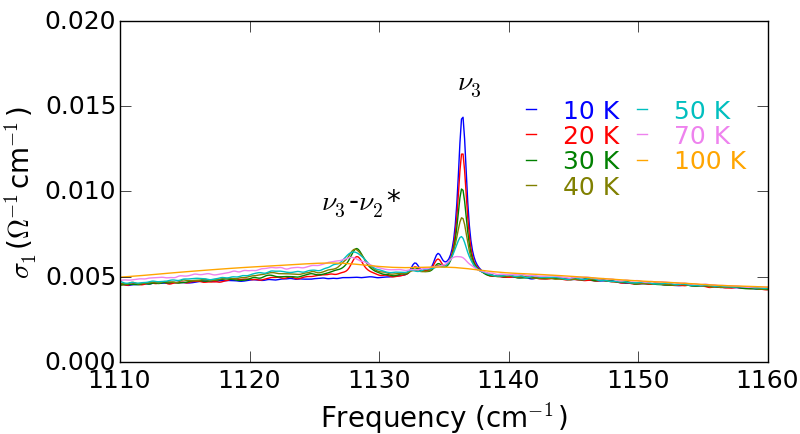}
\caption{\label{v3-si-8552*}$\nu_3$ vibrational modes of Si$_2$O vibrations for sample si-8552*}
\end{figure}
\begin{figure}
\includegraphics[width=0.98\columnwidth]{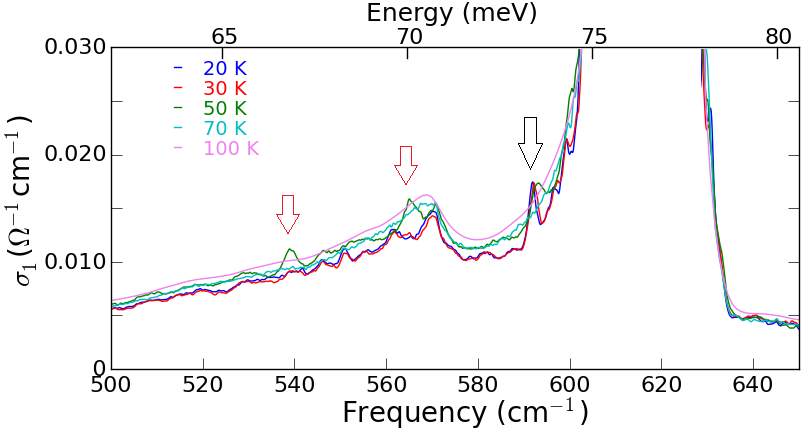}
\caption{\label{ambigu-si-jge}Transition lines in sample si-jge}
\end{figure}

\subsection{Oxygen impurities}

Absorption lines in the very far infrared due to the $\nu_{2}$ mode of vibration of the Si$_{2}$O complex, and modes due to the $\nu_{3}$ vibrational mode in the mid infrared were observed for some samples (see Sec.~\ref{sec-lattice}). The only absorption lines for sample si-8552 (p-type, highest purity) was observed in the mid infrared range. They are due to the $\nu_{3}$ vibrational modes (Fig. \ref{v3-si-8552}). For sample si-st2, in addition to $\nu_3$ vibrational mode in the mid infrared, the $\nu_{2}$ vibrational mode was observed at low temperatures (Fig. \ref{v2-si-st2}). For all other samples, except for si-jge and si-8551, the $\nu_{3}$ vibrational modes were observed in mid infrared ranges. Due to its high O content, the $\nu_{3}$ vibrational mode frequency of sample si-st1 has zero transmission to the extent of measurement accuracy.
Also, temperature sensitive absorption at $\sim$1127.9 cm$^-1$ is seen for samples. This band is almost completely forzen out at the lowest temperatures where $\nu_3$ vibrational mode becomes intense. This behavior is attributed to thermal population of the excited states of the $\nu_2$ mode \cite{bosomworth-1970-oxygen&sivibrat}.

\subsection{Unidentified absorption lines}

Apart from these, we noticed some ambiguous transitions. For sample si-jge, we observe transitions which are similar to thermally populated $1S(E)/1S(T_{2})\rightarrow nP_{0/\pm}$ like transitions at 66.8 meV and also $1S(A_{1})\rightarrow nP_{0/\pm}$ like transition at 73.4 meV (Fig. \ref{ambigu-si-jge}). The latter was also observed for p-type samples si-8548 and si-8550. These lines had transition energy that were higher than those of typical shallow donors or acceptors transition and lower than deep-center transitions \cite{pajot-2010}. However, they are close to shallow Cu acceptor transition in Si (65.8 and 68.7 meV) 
\cite{teklemikel2014cuinsi} and also to $S_{c}X_{3}$ single donor transition in Si (70.454 and 75.769 meV) \cite{pajot-2010}. There is also transition line appearing at 202.7 cm$^{-1}$ (25.14 meV) for samples si-8551 and si-st4 (labeled as * in Figs. \ref{farir-si-st4} and \ref{farir-si-8551}). If this transition is assumed to be of the $1S(A_{1})\rightarrow nP_{0/\pm}$ type due to it being highest in strength at the lowest temperature, and since the position for higher $2P_{0}$ or $2P_{\pm}$ states for donors is pretty well established \cite{faulkner-1969}, then the position of the ground state of this line should be either 36.63 or 31.54 meV which is close to the lithium (Li) ground state energy level in Si (31.24 meV for $1S(A_{1})$ in Si:Li).

\begin{figure}[h]
\includegraphics[width=1\columnwidth]{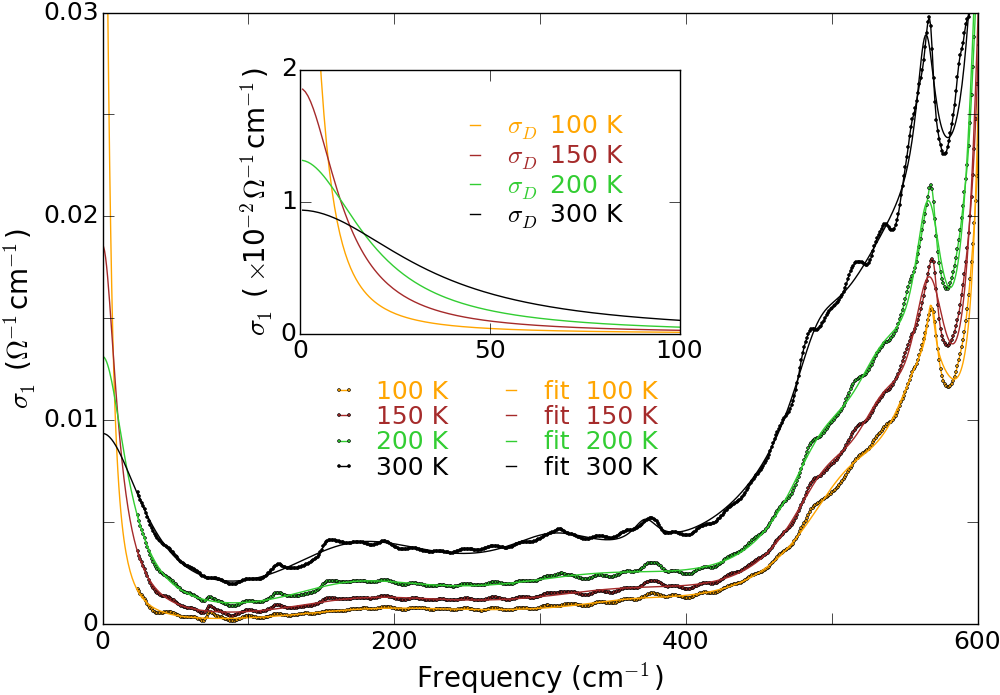}
\caption{\label{sigmafit-si-jge} Optical conductivity of sample si-jge with the fits at various temperatures. Inset: Drude optical conductivity behavior with temperature}
\end{figure}
\begin{figure}[h]
\includegraphics[width=1\columnwidth]{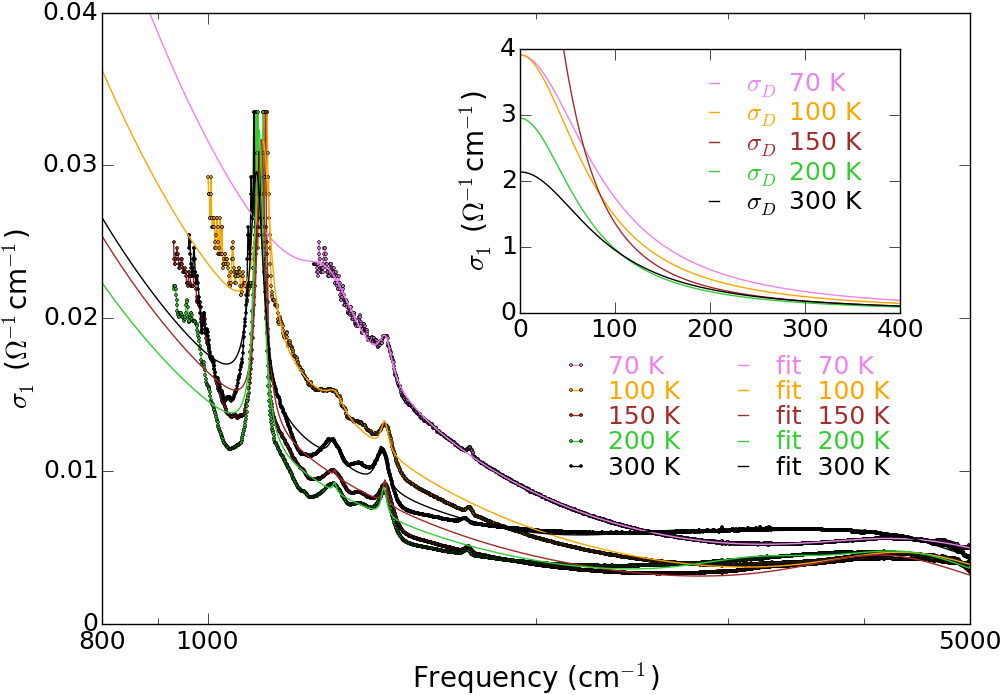}
\caption{\label{sigmafit-si-st1} Optical conductivity of sample si-st1 with the fits at various temperatures. Inset: Drude optical conductivity behavior with temperature}
\end{figure}
\begin{figure}[h]
\includegraphics[width=1\columnwidth]{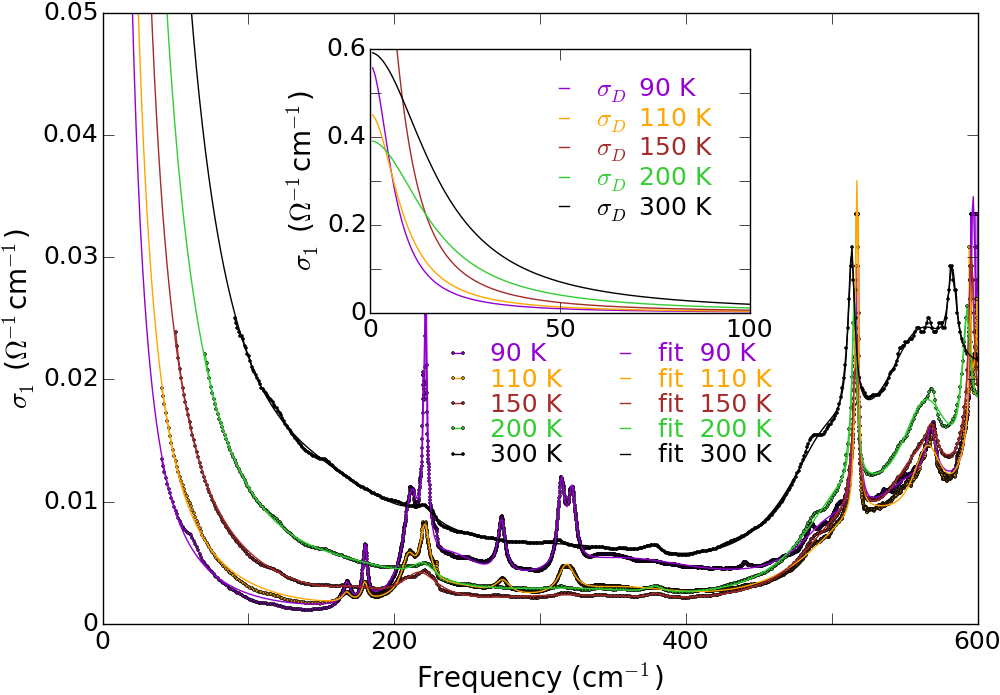}
\caption{\label{sigmafit-si-st2} Optical conductivity of sample si-st2 with the fits at various temperatures. Inset: Drude optical conductivity behavior with temperature}
\end{figure}
\begin{figure}[h]
\includegraphics[width=1\columnwidth]{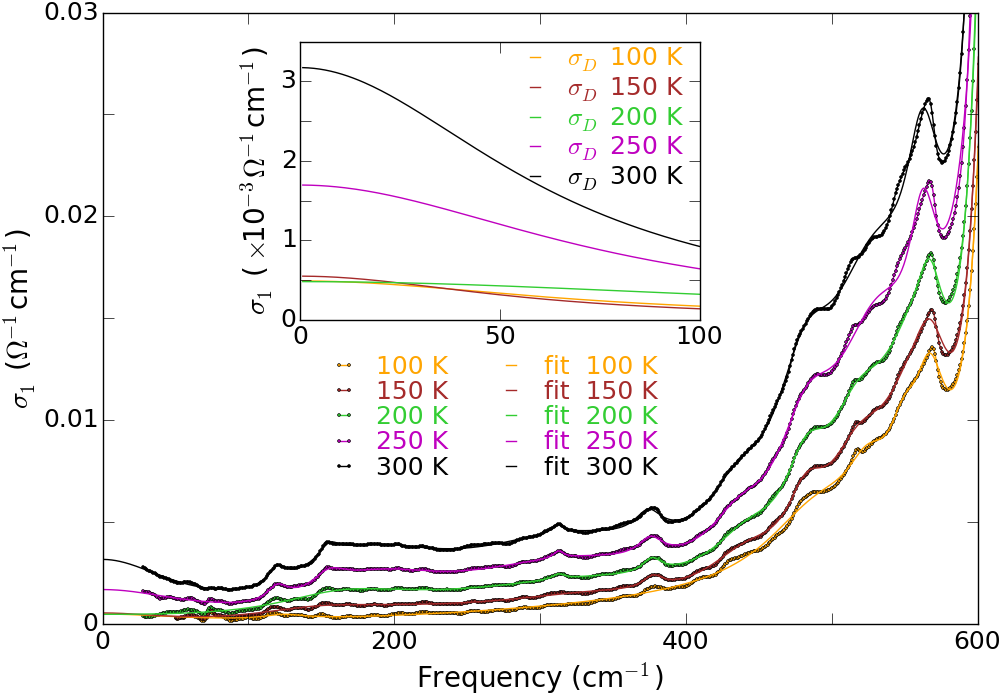}
\caption{\label{sigmafit-si-8547} Optical conductivity of sample si-8547 with the fits at various temperatures. Inset: Drude optical conductivity behavior with temperature}
\end{figure}
\begin{figure}[h]
\includegraphics[width=1\columnwidth]{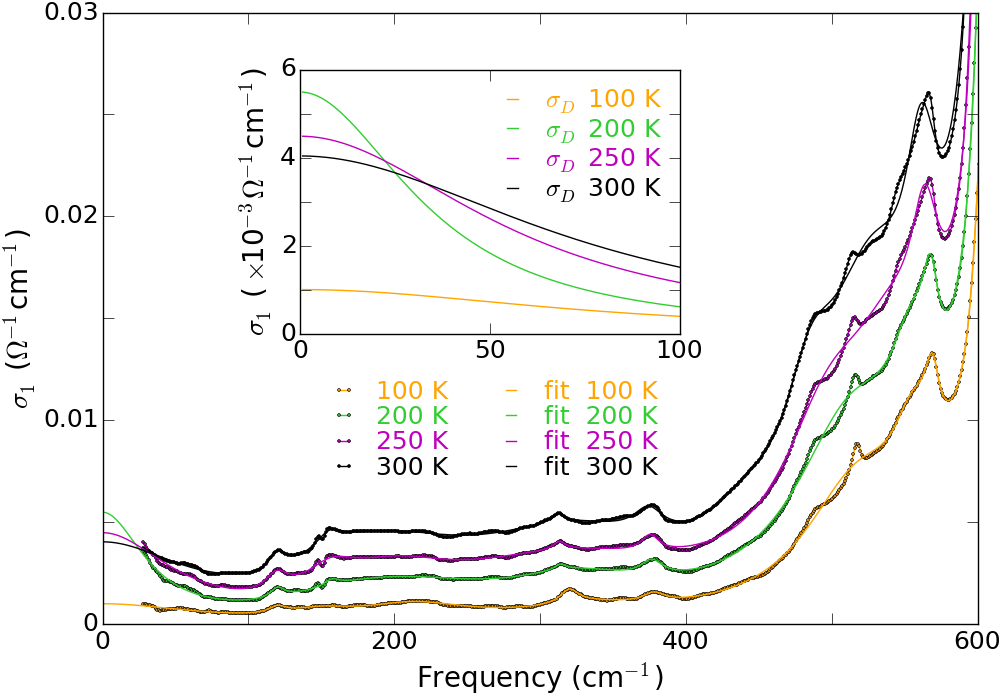}
\caption{\label{sigmafit-si-8548} Optical conductivity of sample si-8548 with the fits at various temperatures. Inset: Drude optical conductivity behavior with temperature}
\end{figure}
\begin{figure}[h]
\includegraphics[width=1\columnwidth]{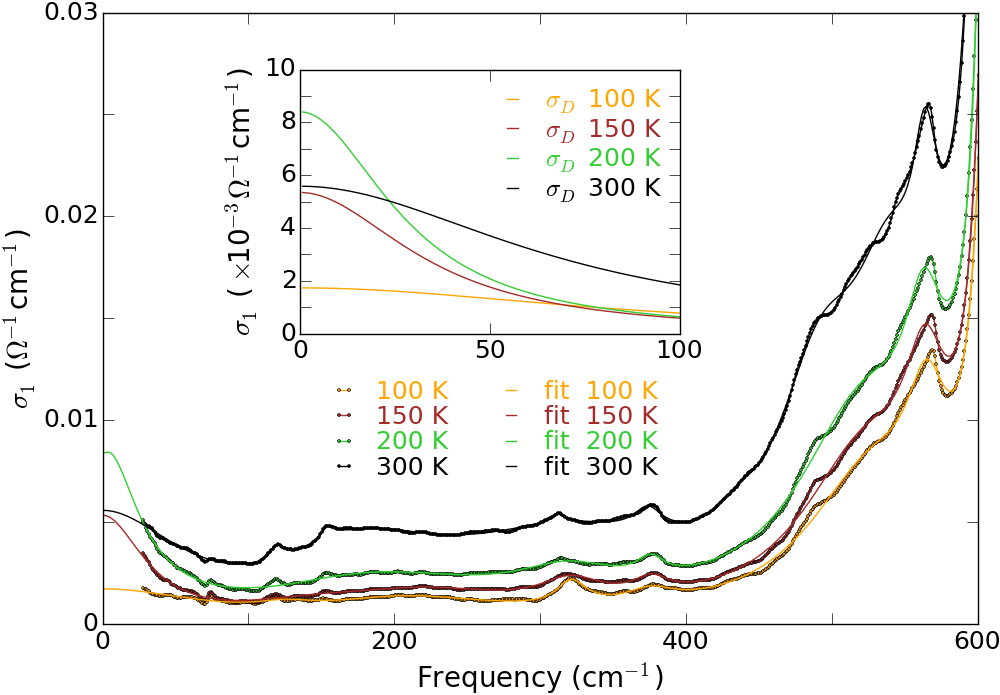}
\caption{\label{sigmafit-si-8550} Optical conductivity of sample si-8550 with the fits at various temperatures. Inset: Drude optical conductivity behavior with temperature}
\end{figure}

\section{Concentration Determination}
\subsection{From the absorption lines $2P_{0}$, $2P_{\pm}$, $3P_{0}$ of P impurity and $1\Gamma_{8}^{-}$, $2\Gamma_{8}^{-}$ and ($1\Gamma_{6}^{-} \oplus 1 \Gamma{_7}^{-} \oplus 4\Gamma_{8}^{-}$) of B impurity}

The observed linewidths, after accounting for instrumental broadening, are limited by several mechanisms \cite{jagannath-1981}, and hence the maximum absorption coefficient $\alpha_{max}$ cannot be taken as a measure to determine the impurity concentration via Beer's law. The absorption line has to be integrated through the spectral extent of the line, around the central frequency of absorption. This process ensures that the instrument's limit on resolution does not affect the concentration determination. As stated in Sec. \ref{sec-beer-lamber}, the integrated absorption $S$ is through Eq. \ref{eq-intg-abs} and concentration could be determined through
\begin{equation}
n_i = K_S^{-1}~S\nonumber
\end{equation}
The absorption lines at $2P_{0}$ ($\sim$275 cm$^{-1}$), $2P_{\pm}$ ($\sim$316 cm$^{-1}$), $3P_{0}$ ($\sim$324 cm$^{-1}$) were used to determine the P dopant concentration. The $1\Gamma_{8}^{-}$ ($\sim$245 cm$^{-1}$), $2\Gamma_{8}^{-}$ ($\sim$278 cm$^{-1}$) and $1\Gamma_{6}^{-} \oplus 1 \Gamma{_7}^{-} \oplus 4\Gamma_{8}^{-}$ ($\sim$320 cm$^{-1}$) lines were used to determine boron dopant concentration. The latter is integrated through spectral extent of three lines because the lines are close together (see Tab. \ref{theory-OS-B}) and cannot be resolved and integrated seperately. The experimentally determined calibration factors are
\begin{equation}
K_S^{-1}=\left\{\begin{array}{c} 
															4.2\times10^{13}~\text{cm}^{-1}  ~\text{for 275 cm$^{-1}$} \\
															1.2\times10^{13}~\text{cm}^{-1}  ~\text{for 316 cm}^{-1} \\
															23\times10^{13}~\text{cm}^{-1}  ~\text{for 324 cm}^{-1} \end{array} \right\}\text{of P}\nonumber
\end{equation}
\begin{equation}
K_S^{-1}=\left\{\begin{array}{lr} 
															6.8\times10^{13}~\text{cm}^{-1}  ~\text{for 245 cm}^{-1} \\
															1.5\times10^{13}~\text{cm}^{-1}  ~\text{for 278 cm}^{-1} \\
															1.7\times10^{13}~\text{cm}^{-1}  ~\text{for 320 cm}^{-1} \end{array} \right\} \text{of B},\nonumber
\end{equation}
where the latter is given for the 320 cm$^{-1}$ band which is the spectral extent of three lines\cite{raey}. For some of the transition lines, $S$ could not be determined because the transmission was zero within the noise and accuracy of the measurements which made shape of the line and $\alpha_{max}$ incorrect. Also, the intensity of some of the lines was too low and blended with noise of the background so no accurate $S$ was determined. The concentration of the impurities calculated through this method ($n_i(S)$) are given in Table \ref{global-table}.

\begin{table*}
\caption{Concentration determined by several methods}
\label{global-table}
\begin{ruledtabular}
\begin{tabular}{c c c c c c}
Sample ID/Type & $\rho$ (vendor)      &  $n_i$ (vendor)    				& $n_i$ ($S$)  					        	 & $n_i$ (OS)         				   & $n_c$ (Drude) \\
               & $\Omega$ cm$^{-1}$     &  cm$^{-3}$    			    & cm$^{-3}$  					         & cm$^{-3}$       						   & cm$^{-3}$ \\
\colrule	

si-jge/n-type    & N/A                & N/A          			    & (8.5$\pm1.3)\times 10^{13}$   & (5.6$\pm1.3)\times 10^{13}$ & (5.2$\pm0.2)\times 10^{13}$\\

si-st1/n-type    & 0.17 							& 3.9$\times 10^{16}$     & -												  & -														& (3.0$\pm0.1)\times 10^{16}$ \\

si-st2/n-type    & 2.8 								& 1.6$\times 10^{15}$     & -												 & -														& (1.7$\pm0.1)\times 10^{15}$\\

si-st4/n-type    & 808 								& 5.2$\times 10^{12}$     & (8.06$\pm1.39)\times 10^{12}$  & (6.1$\pm0.9)\times 10^{12}$ & - 												\\

si-8547/n-type    & 134							 & 3.18$\times 10^{13}$     & -												  & -														& (5.50$\pm0.11)\times 10^{13}$ \\

si-8548/p-type   & 159 							 & 8.37$\times 10^{13}$     & (8.07$\pm0.55)\times 10^{13}$   & (6.63$\pm0.91)\times 10^{13}$ & (8.45$\pm0.14)\times 10^{13}$ \\

si-8549/n-type/    & 2950 						& 1.39$\times 10^{12}$     & (1.04$\pm0.13)\times 10^{12}$  &		-													&			-												\\

si-8550/p-type   & 115 								& 1.16$\times 10^{14}$     & (1.34$\pm0.05)\times 10^{14}$   & (1.06$\pm0.13)\times 10^{14}$ & (1.06$\pm0.01)\times 10^{14}$\\

si-[8551]/[8551*]/n-type    & 10100 		& 4.00$\times 10^{11}$     & [(1.18$\pm0.10)\times 10^{12}$]/[-]  & [(8.0$\pm0.6)\times 10^{12}$]/[-]   &				-		\\

si-[8552]/[8552*]/p-type    & 28500		 & 4.67$\times 10^{11}$     	& -  													& -														&   -												\\
\end{tabular}
\end{ruledtabular}
\end{table*}

\subsection{Concentration from oscillator strengths}

Concentration of the samples were also determined from the oscillator strength calculated by Ref. \onlinecite{clauwsetal-1988-os} and \onlinecite{beinikhes&kogan-1987-os} for P impurity in Si and by Ref. \onlinecite{Buczko&bassani-1992-OSofSi:B} and \onlinecite{pajotetal-1992-OSofSi:B} for B impurity in Si. There, the $S$ of the lines was divided by corresponding theoretical oscillator strength at the lowest temperature achieved. Then the concentration was determined by\cite{andreev-1995}
\begin{equation}
N=\frac{Q*10^{12}~~IA}{f_{th}}
\end{equation}
using $Q$~=~0.987~(Si:P) and $Q$~=~0.907~(Si:B). Here, $f_{th}$ is the calculated values from Table \ref{theory-OS-P} and Table \ref{theory-OS-B}. The concentration was determined form various lines and then averaged. The average and standard deviation of $n_i$ are given in Table \ref{global-table} using this method ($n_i$ (OS)).

\subsection{Concentration from Drude absorption}

\begin{figure}
\includegraphics[width=1\columnwidth]{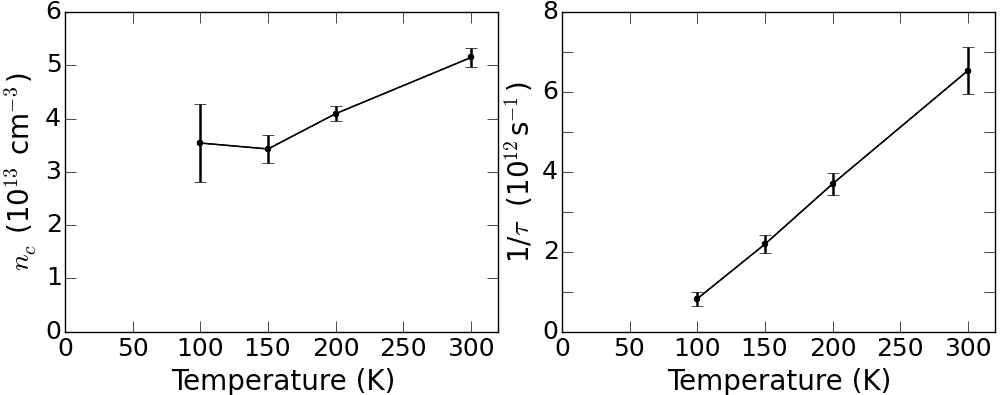}
\caption{\label{error-si-jge}  Carrier concentration and damping constant behavior with temperature for sample si-jge}
\end{figure}

\begin{figure}
\includegraphics[width=1\columnwidth]{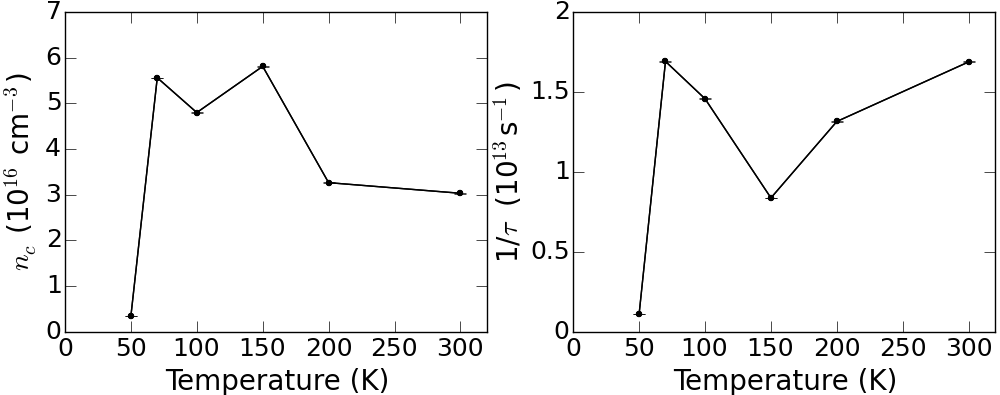}
\caption{\label{error-si-st1}  Carrier concentration and damping constant behavior with temperature for sample si-st1}
\end{figure}

\begin{figure}
\includegraphics[width=1\columnwidth]{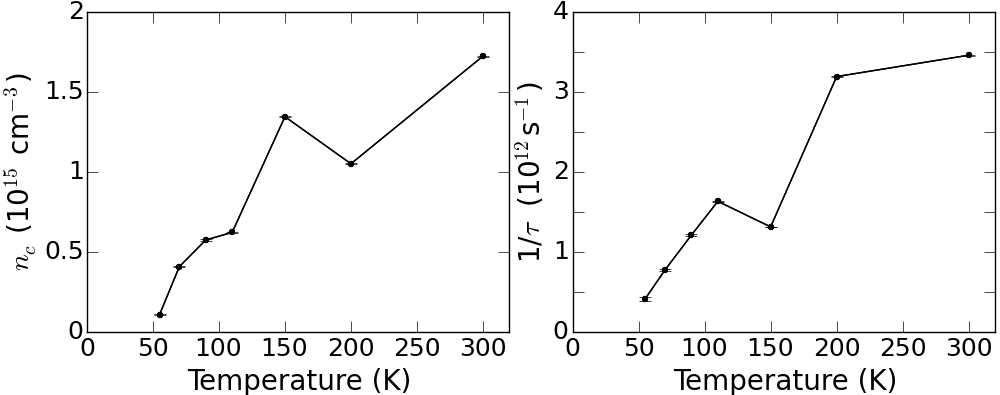}
\caption{\label{error-si-st2}  Carrier concentration and damping constant behavior with temperature for sample si-st2}
\end{figure}

\begin{figure}
\includegraphics[width=1\columnwidth]{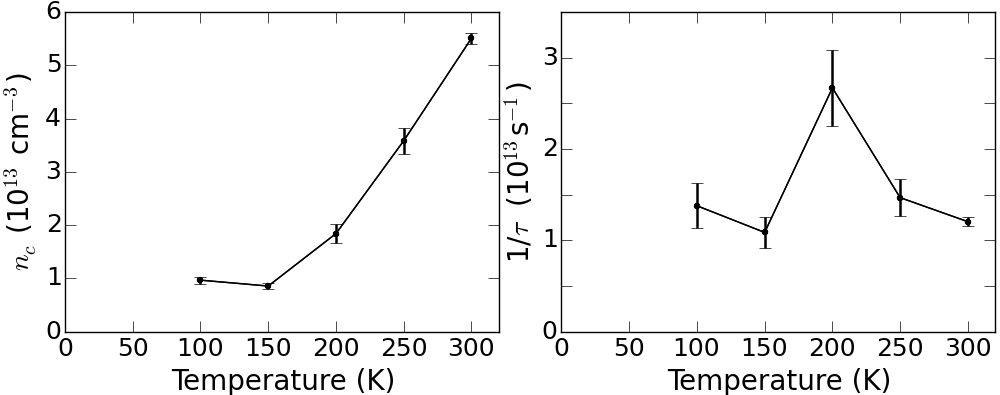}
\caption{\label{error-si-8547}  Carrier concentration and damping constant behavior with temperature for sample si-8547}
\end{figure}

\begin{figure}
\includegraphics[width=1\columnwidth]{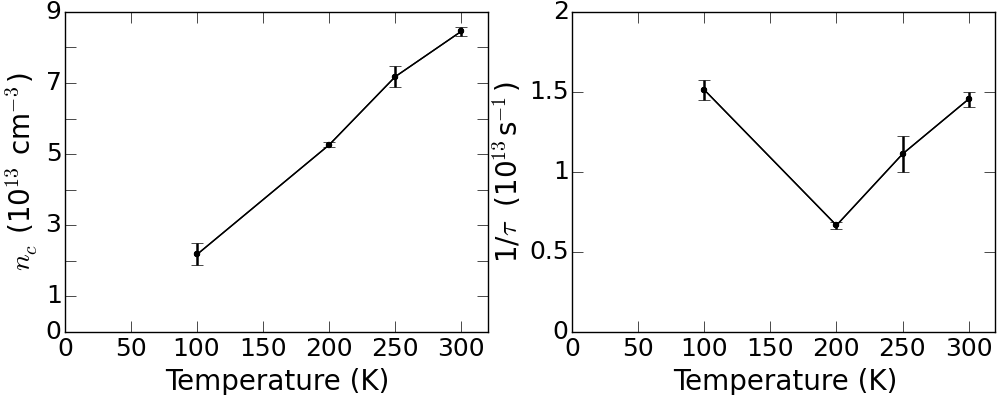}
\caption{\label{error-si-8548}  Carrier concentration and damping constant behavior with temperature for sample si-8548}
\end{figure}

\begin{figure}
\includegraphics[width=1\columnwidth]{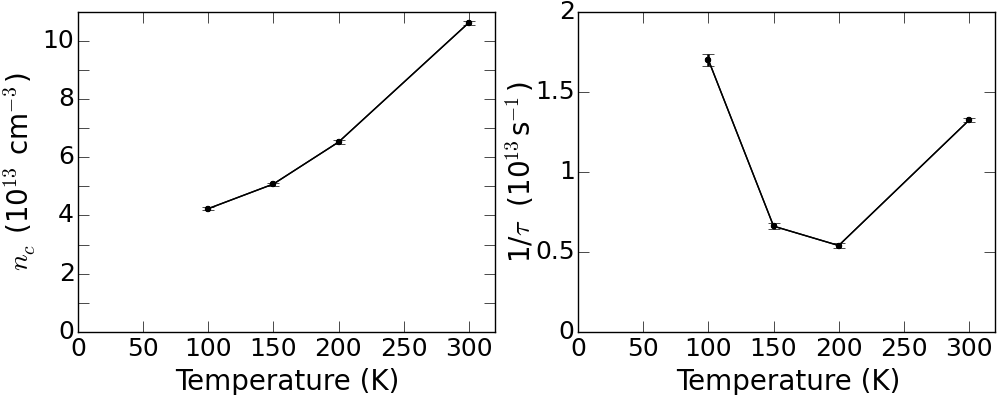}
\caption{\label{error-si-8550}  Carrier concentration and damping constant behavior with temperature for sample si-8550}
\end{figure}

Samples si-jge, si-st1, si-st2, si-8547, si-8548, and si-8550 show Drude like free carrier absorption in the far infrared range of frequencies. If we now consider the fact that the refractive index is relatively constant in this range, we can use sum rule for absorption (Eq. \ref{eq-part-sumrule}) to figure out the carrier concentration\cite{ohbaandikawa-1988-frcar}. The fit to the optical conductivity was done by least square fits \cite{tanner-opteff}. Fits to the optical conductivity are shown in Fig. \ref{sigmafit-si-jge} through Fig. \ref{sigmafit-si-8550} .The insets show the behavior of Drude conductivity with temperature. From the Drude conductivity fit to the 300 K, we deduced the spectral weight which is $\omega_{p}$, central frequency ($\omega_{0}=0$ for Drude model), and a damping constant $\gamma=1/\tau$. From the formula for the plasma frequency (Eq. \ref{eq-sumrule}), we determined the concentration for the samples above using corresponding conductivity effective masses (Sec. \ref{sec-OS}). Tab. \ref{global-table} shows corresponding carrier concentrations ($n_i$ (Drude)) found from plasma frequency. The behavior of the carrier concentration and the damping constant are shown in Fig. \ref{error-si-jge} through Fig. \ref{error-si-8550}. The carriers freeze out as the temperature is decreased, which means that donor or acceptor centers become neutralized.
It is important to note here that sample si-8547 shows Drude carrier absorption, but does not show any sharp absorption lines at low temperatures except for $v_3$ vibrational mode of Si$_2$O. This could be attributed to the fact that oxygen in Si can become a thermal donor, which is an electrically active center, and contribute to free carrier absorption in far infrared spectra \cite{michel&kimmerling-1994-elec-O-in-Si}.

\section{Conclusion}

Most of our samples displayed a good optical transmission in infrared ranges which is a good sign for the immersion gratings. As we have previously noted, the high refractive index helps to reduce the instrument's size or increase the resolution of a high resolution infrared spectrometers utilized in astronomy \cite{ge-2014-berik, ge-2012}. However, it is clear that the materials are much affected by the multiphonon absorption in the mid infrared and impurity absorption in far infrared. Therefore, these absorption ranges may not be helpful in identification and characterization of extrasolar planets. In addition, if Si has high oxygen content, the far infrared ranges of 10 - 30 cm$^{-1}$ are not as feasible too due to $v_{2}$ vibrational mode of Si$_{2}$O.

Concentrations shown in Tab. \ref{global-table} are net impurity concentration and they do not account for compensated impurities. Samples  si-8551*, si-8552 and si-8552* did not display any free carrier or impurity absorption so we could not determine the concentration. We attribute some discrepancy with the vendor concentration, which was determined from the vendor resistivity, to the fact that concentration obtained from transmission method is the average over the thickness of the samples, while the four-point probe method gives surface resistance which is then converted to the bulk resistivity values. Also, the resistivity measurements require an Ohmic contact which is hard to achieve with highly resistive polished samples.

Overall, the transparency in the near infrared is pretty high and the Si is readily usable at that range of frequencies for astronomy. The samples displayed a good transmittance in the near infrared as long as the net impurity concentration was below $\sim 10^{15}$ cm$^{-3}$. So SIGs could be constructed and used in frequency range of $\sim$ 1500 - 8400 cm$^{-1}$ ($\lambda\approx$ 1.2 - 6.6 $\mu$m) at room temperature. At liquid helium temperature (namely 4.2 K), due to band gap expansion, the high frequency limit can be extended to $\sim$ 9500 cm$^{-1}$ ($\lambda\approx$ 1.05 $\mu$m). It is obvious that for SIGs one needs to have as pure samples as possible but going below $\sim 10^{15}$ cm$^{-3}$ seems to be just enough for the near infrared regions.

For the midinfrared region, one is limited by the multiphonon absorption of the Si. Midinfrared multiphonon absorption appears in frequency range of $\sim$ 450 - 1500 cm$^{-1}$ ($\lambda\approx$ 6.6 - 22.2 $\mu$m), and it lessens a bit at low temperatures. However, it is still high and there is no good way of utilizing this region except when the multiphonon absorption is not as drastic.
 
In the far infrared, operating SIGs at room temperature will be limited by the far infrared multiphonon absorption that involves acoustical branch. However, it softens out as the temperature is decreased and SIGs could be utilized in the far infrared region which is in frequency range of $\sim$ 10 - 450 cm$^{-1} (\lambda\approx$  22.2 - 1000 $\mu$m). Yet, the Si has to be more pure than $\sim 10^{12}$ cm$^{-3}$  if to be utilized in these regions. This purity will ensure that the impurities will not affect the performance of SIGs. If not so pure samples are available, then at low temperatures, residual impurity absorptions dominate and some regions of far infrared are not feasible. 
One might want to avoid the $v_{2}$ vibrational mode of Si$_{2}$O, phosphorus and boron absorption lines to operate the instrument. For n-type samples, the useful range of frequencies would be $\leq$ 275 cm$^{-1}$ ($\lambda\geq$ 36.4 $\mu$m) which is the lowest transition of the P impurity in Si (i.e., $1S\left(A_{1}\right)\rightarrow2P_{0}$ absorption line of Si:P). And also if it has too much oxygen content, then the range below 30 - 50 cm$^{-1}$ is not utilizable because the v$_{2}$  vibrational modes of Si$_{2}$O happen in those regions. For the p-type samples, considering the boron impurity, the utilizing range would be $\leq$ 245 cm$^{-1} (\lambda\geq$ 40.8 $\mu$m) due to the ground state to excited state transition (i.e., $1\Gamma_{8}^{+}\rightarrow n\Gamma_{n}^{-}$ absorption line in Si:B) \cite{ramdas&rodriguez-1981-bigrev}. There might be some other impurities which might affect the far infrared transmittance of Si, but these usually are not as significant as P and B impurities in Si. 

The behavior of impurities in Si is one that was expected for shallow impurities. At low temperatures we see hydrogenic like transitions and at intermediate temperatures we observe transitions from valley orbit split ground state. Concentrations were determined from $2P_{0}$, $2P_{\pm}$, $3P_{0}$ for n-type and $1\Gamma_{8}^{-}$, $2\Gamma_{8}^{-}$ and ($1\Gamma_{6}^{-} \oplus 1 \Gamma{_7}^{-} \oplus 4\Gamma_{8}^{-}$) for p-type absorption lines, oscillator strengths at low temperatures and Drude absorption from room temperature. The nature of some of the lines were ambiguous and have yet to be understood at which stage of production they have appeared. Overall, the host material of SIGs need to be of high purity in order to give adequate performances which is going to be better (for transmittance will increase) for lower temperatures.

\bibliography{silicon2}

\end{document}